\documentclass{article}
\usepackage{lipsum}
\usepackage[sorting = none]{biblatex}
\usepackage{array}
\usepackage{url} 
\usepackage{hyperref}
\usepackage{multirow}
\usepackage{tabularx}
\usepackage{rotating}
\usepackage{amsmath}
\usepackage{amssymb}
\usepackage{geometry}
\usepackage{graphicx}
\usepackage{gensymb}
\usepackage{graphics}
\usepackage{xcolor}
\usepackage{amsmath}
\usepackage{amsfonts}
\usepackage[t,lf]{spectral}
\usepackage[T1]{fontenc}
\usepackage[center]{caption}
\usepackage{adjustbox}
\usepackage{changepage}
\usepackage[utf8]{inputenc}
\usepackage{longtable}
\usepackage{subcaption}
\usepackage{longtable}
\usepackage{lscape}
\usepackage{subfiles} 
\usepackage{seqsplit}
\usepackage{array}
\usepackage{wrapfig}
\usepackage{xcolor}
\usepackage{comment}
\usepackage{tabularx}
\usepackage{tikz}
\usetikzlibrary{trees,positioning,shapes,shadows,arrows}
\usepackage{pgf-pie}
\usepackage{pgfplots}
\usepackage{lscape}
\usepackage{pgfplotstable}
\usepackage{multirow}
\usepackage{mathtools}
\usepackage{comment}
\usepackage{setspace}
\usepackage{algorithm}
\usepackage{algpseudocode}

\providecommand{\keywords}[1]{\textbf{\textit{Index terms---}} #1}

\graphicspath{ {images/} }
\usepackage{supertabular,booktabs}

\begin{document} 

\title{A survey on scheduling and mapping techniques in 3D Network-on-chip}

\author{
  Simran Preet Kaur, Manojit Ghose, Ananya Pathak, Rutuja Patole \\
  Department of Computer Science and Engineering \\
  Indian Institute of Information Technology Guwahati \\
  India \\
  \texttt{\{simran.kaur21, manojit, ananya.pathak, rutuja.patole\}iiitg.ac.in} \\
  }

\begin{abstract}
Network-on-Chips (NoCs) have been widely employed in the design of multiprocessor system-on-chips (MPSoCs) as a scalable communication solution. NoCs enable communications between on-chip Intellectual Property (IP) cores and allow those cores to achieve higher performance by outsourcing their communication tasks. Mapping and Scheduling methodologies are key elements in assigning application tasks, allocating the tasks to the IPs, and organising communication among them to achieve some specified objectives. The goal of this paper is to present a detailed state-of-the-art of research in the field of mapping and scheduling of applications on 3D NoC, classifying the works based on several dimensions and giving some potential research directions.
\end{abstract}


\maketitle

\setstretch{1.0}

\section{Introduction}

Nowadays, technological innovation allows us to seamlessly implement increasingly complex applications on computing systems equipped with powerful processors. Each processor generation outperforms the previous one, which entails incorporating more logic onto the silicon \cite{multipr}. But there are two issues with this approach. The first is that we are losing the ability to downsize transistors and the logic and memory blocks that are incorporated in them \cite{trans}. The second issue is that individual chips have reached their maximum size. Thus new technologies bring the concept of hybrid bonding and integrate an entire system on a single chip, known as System-on-Chip (SoC). 

To address the increasing application demands, SoC architecture has become more advanced, evolving from a single processor to a multiprocessor system-on-Chip (MPSoC) \cite{book}. This progress was made possible by the shrinkage of transistors, which allowed for the integration of billions on a single chip \cite{2006_Moore}. MPSoCs may now accommodate tens of Intellectual Properties (IPs), such as processor cores, memory modules, and other I/O components \cite{mpsoc}. These systems boost performance by combining many homogeneous or heterogeneous processors, connected together to form a network.

\begin{figure}
  \includegraphics[width=\linewidth]{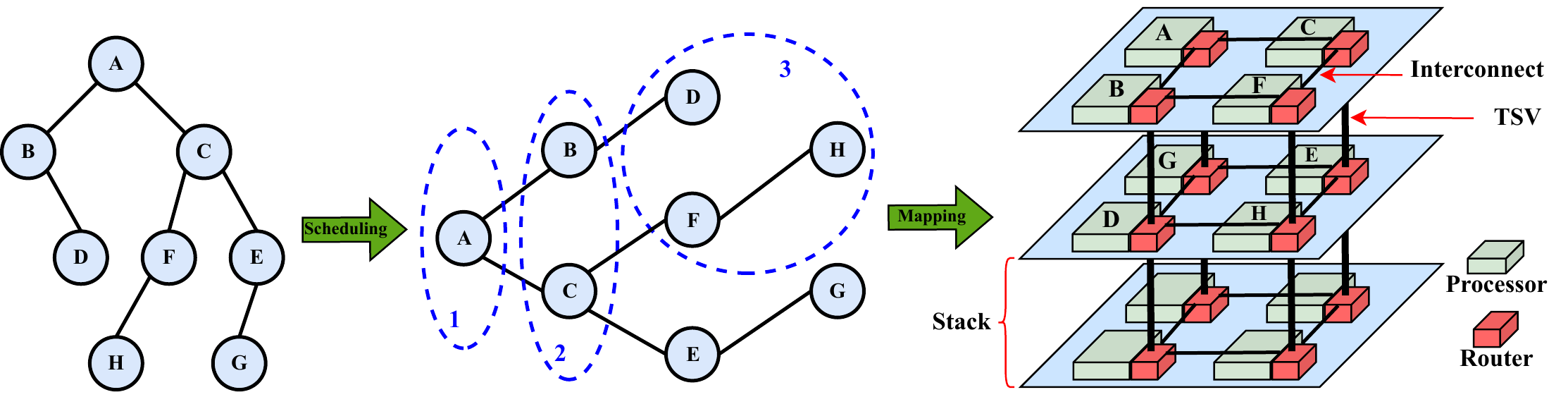}
  \caption{Scheduling and Mapping in 3D NoC}
  \label{figMapSch}
\end{figure}

The Network-on-Chip (NoC) paradigm has evolved as an enabler for the integration of numerous embedded cores on a single device. NoC has been presented as a way to replace traditional interconnections in SoC and MPSoC-based buses, guaranteeing flexible communication and high necessary bandwidth \cite{noc}. Because of its regularity and simplicity, the 2D-mesh topology is the most researched and utilised NoC topology. However, when the number of cores rises, resulting in higher network size, network performance might suffer drastically\cite{tatas}. The adoption of 3D NoC is the proposed method to increase performance in terms of power savings and average hop-count \cite{topo, feero}. A 3D NoC-based MPSoC is built by stacking numerous levels on top of one other and connecting them vertically via through-silicon vias (TSVs). Since 3D NoC-based MPSoCs can integrate more processors and execute more programs, establishing efficient and accurate mapping and scheduling techniques is necessary. Previous surveys cover various mapping techniques aiming 2D architectures \cite{Carvalho, ft}, while authors in \cite{survey_security} review security threats and countermeasures targeting regular 3D NoC. However, to the best of our knowledge, task mapping and scheduling in 3D NoC are still uncharted areas. Mapping and Scheduling steps follow 3D NoC specialization and their job is to implement the given application into the selected architecture, which primarily means to assign and order the tasks and communications of the application into the resources of the architecture such that the design goals are optimized.
Figure \ref{figMapSch} demonstrates the scheduling and mapping of an application onto a 3D NoC system. The application is represented as a Directed Acyclic Graph (DAG) where each node indicates some tasks (computation unit). Each task has a weight associated with it that represents its cost of execution. The process is achieved using two steps. At first, the tasks of the DAG are ordered for their execution. This is termed as scheduling. In the figure, the tasks are grouped based on their execution order (numbered as 1, 2 and 3). The next step is the mapping where the tasks are to be allocated to the cores.\\
Area and power consumption for 3D NoC can be accurately estimated  using the findings of latency, throughput, and communication bandwidth obtained from the simulated network. After designing the communication infrastructure, communication technique, and evaluation framework for NoC, the final and most crucial process is to associate and determine the appropriate arrangement and placement of application activities on different cores. This is highly beneficial in reducing 3D NoC's communication costs and overall performance, and it constitutes the fourth dimension, mapping and scheduling in 3D NoC \cite{map1},\cite{map2},\cite{sched1}. We concentrated primarily on this research topic in this study to assess mapping and scheduling methodologies in 3D NoC architectures.
\begin{figure}[bt]
  \includegraphics[width=\linewidth]{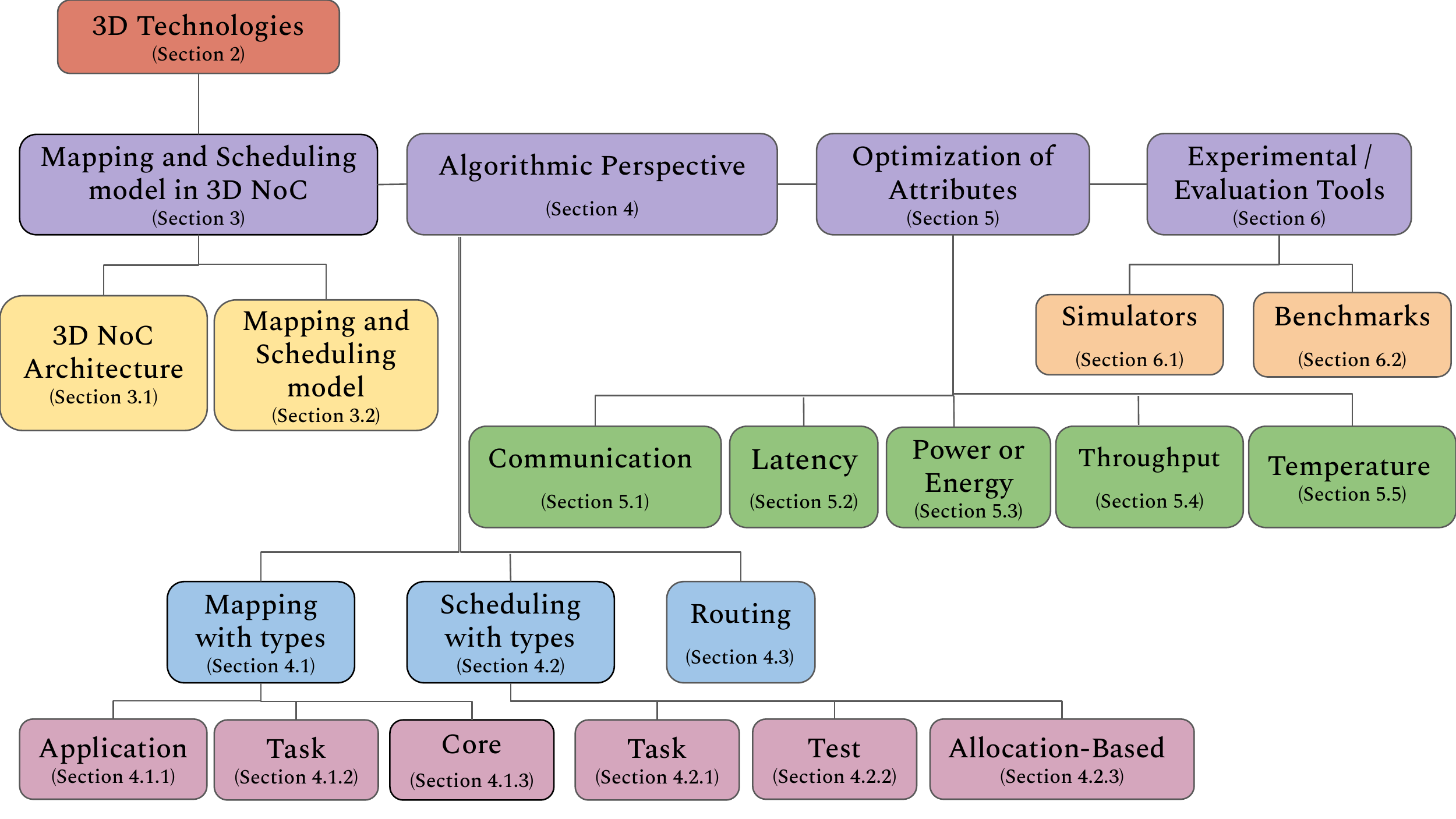}
  \caption{Paper organization}
  \label{Organization}
\end{figure}
The aim of this paper is to provide a comprehensive overview of all the aspects of mapping and scheduling (task graph generation, scheduling, optimization techniques, simulation setup, and performance evaluation metrics). The taxonomy we propose for classifying all the papers given in this work is shown in Figure \ref{Organization} hierarchically. Our work illustrates mapping methodologies using different categorization than \cite{survey_multicore}, \cite{noc_survey} or \cite{scheduling_survey}. In addition, we classify the strategies according to certain optimization metrics. The organization of this survey is as follows: first, we provide the reader with the need and an overview of the mapping and scheduling model in 3D NoC, presented in Section \ref{sec3DMappingAndSched}. We present a brief overview of the underlying technologies of the 3D NoC in Section \ref{sec3DTechnology}. In Section \ref{AlgorithmsMapSched}, the related work of mapping and scheduling techniques in 3D NoC architectures is discussed. Section \ref{Attributes} talks about the optimization of various attributes followed by Section \ref{Experiment} where the experimental details are provided and Section \ref{Conclusions} concludes this paper by giving insight into future opportunities. 

\section{3D Technologies} \label{sec3DTechnology}
We need shorter and denser connections to transfer enormous volumes of data on the same chip, which can only be accomplished by stacking one chip on top of another. In a 3D design, connecting two chips face to face means making hundreds or thousands of micrometer-long connections per square millimeter. These small, dense connections allow data to flow from one piece of silicon to another as quickly and efficiently as if the two were on one chip. Microbumps, 3D integration or merging 2.5D and 3D technology via Through Silicon-Vias (TSVs) contribute to certain significant ways for establishing dense connections.

\emph{1) 2.5D integration with 3D:} In 2.5D structure, there is no stacking of dies on dies, but dies are on Silicon Interposer. The dies are packed into a single package in a single plan and both are flip-chipped on a silicon interposer. Interposers and dies are layered one on top of the other in a 3D framework using TSVs. Intel employs 2.5D technology in combination with 3D by integrating two 3D stacks of chiplets with high-density interconnects \cite{Ponte}. The dies also link high-bandwidth memory and an I/O chiplet to the biggest chiplet (base tile), on which the rest of the chips are placed. The base tile subsequently uses 3D stacking technology to stack compute and cache chiplets atop it. The technology creates a dense matrix of vertical die-to-die connections between two chips.

\emph{2) 3D integration:} It refers to stacking of IC chips and connecting them using Microbumps or TSVs, and stacking multiple device layers on a single chip, which may or may not use very-fine-pitch TSVs to form the interconnect. Even if one of the chips in the stack lacks a single transistor, 3D integration is believed to speed up calculations.
\emph{Microbumps} are a miniature version of flip-chip packaging technology where solder bumps are added to the endpoints of the interconnects at the top (face) of a chip. The chip is then flipped onto a matching set of interconnects on a package substrate, and the solder is melted to make a connection. To stack two chips using this approach, the first chip must have small copper pillars protruding from the surface. These are then coated with a solder microbump, and the two chips are linked face-to-face by melting the solder.
\emph{ TSVs} are copper interconnects that run vertically down the silicon of a chip. They do not span the whole depth of a wafer, hence the backside of the silicon has to be ground away until the TSV is exposed. In 3D stacked chips, the chips are bonded together such that their interconnects are face-to-face. So, the TSVs supply the stack with power and data access.
Graphcore implements 3D integration by connecting a power-delivery chip to its AI processor \cite{graphcore}. The power-management die is densely packed with capacitors and TSVs.

\emph{3) Hybrid bonding: Copper bonding:} Hybrid Bonding connects copper pads at the top of a chip's interconnect stack to copper pads on another chip. The pads in hybrid bonding are in small recesses surrounded by an oxide insulator. When the copper pads of two chips are placed opposite to each other at normal temperature, the insulator is chemically activated and bonds instantly. The copper pads expand in an annealing stage and bridge the gap to generate a low-impedance link. It offers significant bandwidth and power benefits over state-of-the-art microbump-based approaches. AMD's Zen 3 with 3D V-Cache \cite{V3Cache} is a 3D stacked device that uses hybrid bonding to attach additional cache to a high-performance processor.

\section{3D NoC Mapping and Scheduling Model} \label{sec3DMappingAndSched}
\subsection {3D NoC Architecture}

\begin{figure}
\centering
  \includegraphics[scale = 0.70]{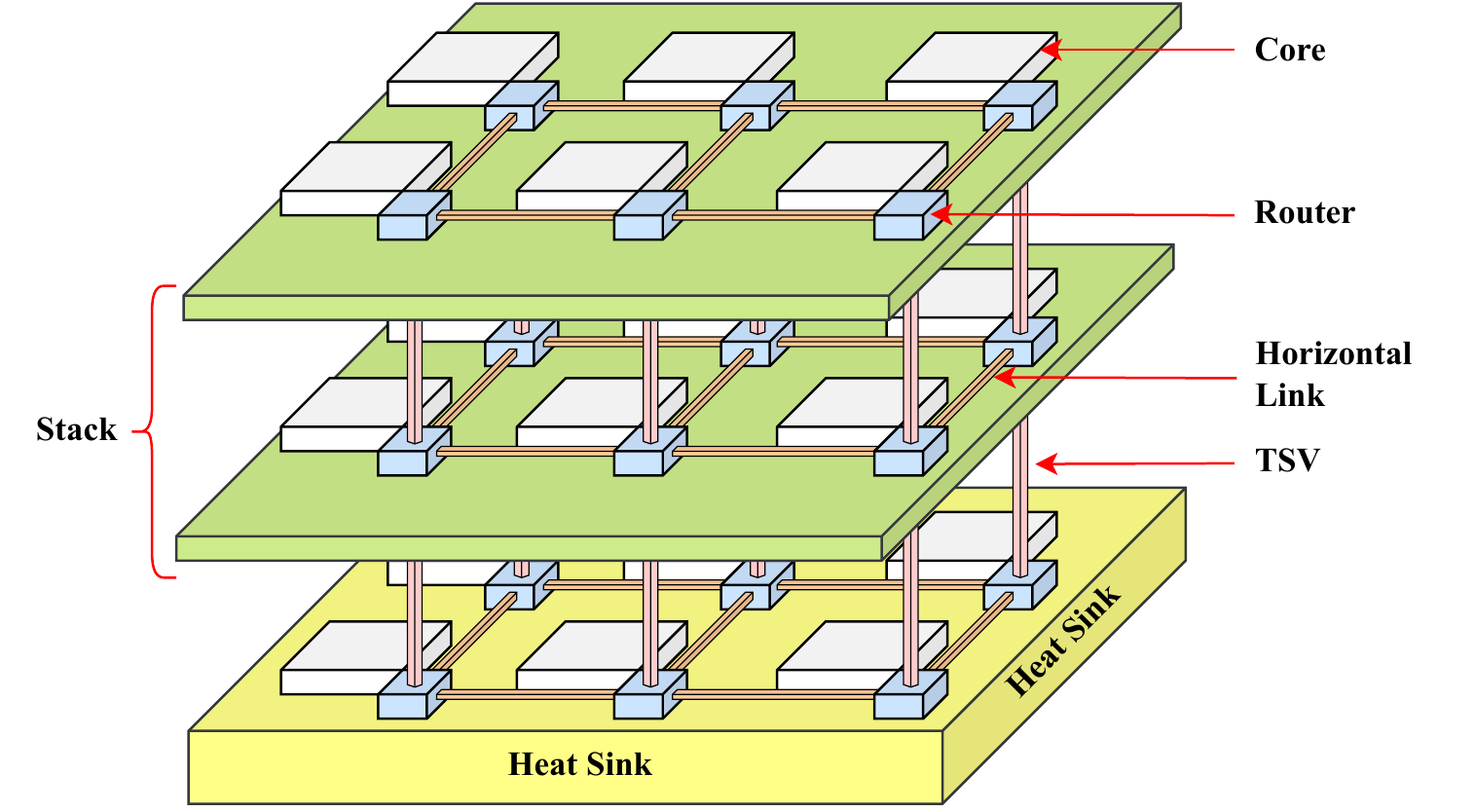}
  \caption{A 3D NoC architecture}
  \label{figArchitecture}
\end{figure}

3D-MPSoC is represented as a regular mesh of tiles built across many layers and linked via Network-on-Chip (NoC). It can be seen in Figure \ref{figArchitecture}. Every tile has a processing core that is connected to a network router. Routers are connected together through links, which constitute the overall architecture.
Packet-mode communication is employed in the architecture, hence all the data transferred in the network is loaded into packets, and all packets are split down into small pieces known as flits. Each core block is further subdivided into compute blocks and L2 cache \cite{3dnoc}. Lateral communication is carried out via NoC, while vertical links are implemented using TSVs.

3-D NoC architecture is impacted by three factors: 1) Topologies, 2) Router designs and 3) 3-D NoC Design methodology

\emph{1) Topology} determines how the nodes in the network are connected with each other \cite{top1}. 
In a multiple-hop topology, packets may travel via one or more intermediary nodes before reaching the destination node. Mesh and torus topologies are common multiple-hop topologies used in 3D NoCs. The authors in \cite{topology} present a three-layer NoC with a unique topology for each layer to fulfill its individual cost-performance requirements, and routers are connected by crossbar switches in the same pillar but at separate layers. In terms of network performance and energy dissipation, Feero \emph{et al.} \cite{feero} compared and examined 3-D mesh-based topologies (symmetric 3-D mesh, stacked mesh, and ciliated mesh) and 3-D tree-based architectures.

\emph{2) Router Designs:} Scalable NoC contains a router at each hop to control access to shared resources \cite{router}. The following is how a conventional NoC router operates. The router first writes a received packet into a buffer and computes the packet's admissible output port at the same time. The Virtual-Channel(VC) allocation mechanism then assigns the packet a free virtual channel of the output channel. When a VC is obtained successfully, the packet arbitrates for accessing the crossbar ports. If everything goes well, the packet can proceed to a neighbor router via the switch traversal(ST) and link traversal(LT) stages. Going through these stages sequentially generates significant delay on each hop, making communication inefficient. Furthermore, if any of the desired resources (e.g., buffer, crossbar) are not available, a packet is stalled, increasing the packet delay. To enhance the performance and reduce the manufacturing cost of 3D NoCs with minimal distortion to the modularity, inhomogeneous architectures have been proposed to merge 2D and 3D routers in 3D NoCs \cite{homogenous}. However, inhomogeneous 3D NoCs, have a performance trade-off due to the restricted number of 3D routers. So, authors in \cite{52} proposed a co-design of routing algorithms and router architectures to mitigate the limitations of NoCs in heterogeneous 3D SoC. Another efficient three-stage pipelined adaptive VC router is proposed by combining the low-complexity bypassing technique with adaptive routing to balance the traffic in hybrid NoCs to achieve low-latency communication under various traffic loads \cite{51}.

\emph{3) Design Methodologies:} Various designs for the efficient 3D NoC architecture have been presented by modifying or merging topologies, routers, or vertical links. In 2D NoC, data packets are transferred between distant cores, where the number of hops increases the delay substantially. To address the issue of interconnect delay, 3-D integration is adopted. TSV is a well-known 3D integration method but they occupy a large area on the chip and also lead to cost increase due to additional fabrication processes and low yield. To tackle these issues, wireless interconnect using capacity coupling and inductive coupling is an excellent substitute for TSVs. 
Joseph \emph{et al.} \cite{65} use inductive coupling for application-specific 3D NoC design that reduces the design complexity caused by channel interference while increasing channel efficiency.
Next, in order to improve the overall system power, performance, and reliability of these mesh architectures, we also need to address routing issues and buffer utilization. Rahmani \emph{et al.} in \cite{36} introduce an architecture that has a higher tolerance for single-bus failure compared to existing architectures.
Dynamic power decreases quadratically as supply voltage decreases, hence using several supply voltages is a well-known low-power solution. Siozios \emph{et al.} \cite{17} show that a single power supply for designing homogeneous  NoCs is inefficient and introduces a 3D NoC architecture with multiple supply voltages. In addition, a high-level mapping algorithm for facilitating application mapping onto such NoCs is presented where the applications are clustered into groups and these groups are mapped onto the proposed architecture.

Performance and energy consumption are prime concerns in today’s Systems-On-Chip (SoCs) due to the continuous scaling of process technology. 
Scaling permits design engineers to integrate as much IP as possible on a chip in order to conduct multitasking and meet client requirements. However, it leads to reduced interconnect performance and scaling issues have become a constraint for multiprocessor system-on-chip communication. As a result, authors in \cite{89} propose an effective design and scaling of on-chip interconnect architectures which is divided into four steps. The initial stage is to identify the IP cores and their size, as well as the core's probable mapping with regard to the router. The core size is equal to the total number of IP cores present in the MPSoCs after clustering, which will be mapped onto the topology. The number of mapping possibilities is equal to the number of routers in the topology architecture, which determines the number of generations. The second step is to estimate the fitness function, which yields the total energy consumption(EC) and communication costs(CC) for mapping the IP cores onto routers. The third stage determines the \emph{local best} core and \emph{global best} core. Every core has a local best, which is the permutation of core locations with the lowest EC and CC values among all permutations of that core. For a certain transmission, the IP cores with the lowest network energy and CC values are the global best of that transmission. The final stage is the evolution over many generations where a swap operator and a swap sequence are introduced through which the \emph{local best} and \emph{global best} are updated, which contribute to the discovery of the optimal solution for the lowest network energy and CC based mapping.
Extension of NoC architecture to three dimensions, brings benefits of both approaches together, meaning more performance of communication, better scalability, and lower power consumption. The last one is due to the shorter wire length and interconnects capacitance. However, despite all these benefits, a critical dilemma is intensified at higher integration levels. As device density increases, power density increases too and as a result, thermal management and required cooling solutions become more challenging. To address the aforementioned constraints, we explore several mapping and scheduling techniques for regular 3D NoC-based MPSoCs.

\subsection{Mapping and Scheduling Model} 
Mapping and scheduling may be performed both online and offline. Offline or static mapping and scheduling are conducted prior to application run-time. Since static mapping and scheduling are executed once during compilation, they have no effect on application performance at run-time. Online or dynamic mapping and scheduling entail the assignment and sorting of tasks while the application is running. This should result in a better solution, but the computational complexity of mapping and scheduling techniques may increase the application's latency and energy consumption at run-time. 

The primary focus of our research here is to provide details about scheduling and mapping techniques in 3D NoC systems. While doing so, we also need to talk about routing to some extent because routing significantly impacts the effectiveness of scheduling and mapping. Since communications between cores in a regular NoC are achieved by routing messages hop-by-hop from the source to the destination, it is essential to include routing algorithms in the discussion of mapping and scheduling for the effectiveness of 3D NoC architectures. Thus, along with considering papers on scheduling and mapping, we also covered some routing strategies. This is shown in Figure \ref{figClassification}.


\tikzset{
  basic/.style  = {draw, text width=5cm, drop shadow, font=\sffamily, rectangle},
  root/.style   = {basic, rounded corners=2pt, thin, align=center, fill=blue!20},
  level 2/.style = {basic, rounded corners=6pt, thin,align=center, fill=pink!60, text width=3cm},
  level 3/.style = {basic, thin, align=left, text width=7em, fill=green!20}
}
\begin{figure}
    \centering
\begin{tikzpicture}[scale = 0.88,
  level 1/.style={sibling distance=11.8em, level distance=5em},
  edge from parent/.style={->,solid,black,thick,sloped,draw}, 
  edge from parent path={(\tikzparentnode.south) -- (\tikzchildnode.north)},
  >=latex, node distance=1.3cm,edge from parent fork down]
  
\node[root] {\textbf{Considered Papers}}
  child {node[level 2] (c1) {\textbf{Mapping}}}
  child {node[level 2] (c2) {\textbf{Scheduling}}}
  child {node[level 2] (c3) {\textbf{Routing}}};
  
\begin{scope}[every node/.style={level 3}]
\node [below of = c1, xshift=15pt] (c11) {General Mapping};
\node [below of = c11] (c12) {Application Mapping};
\node [below of = c12] (c13) {Task Mapping};
\node [below of = c13] (c14) {Core Mapping};
\node [below of = c2, xshift=15pt] (c21) {General Scheduling};
\node [below of = c21] (c22) {Task Scheduling};
\node [below of = c22] (c23) {Test Scheduling};
\node [below of = c23] (c24) {Allocation Based Scheduling};
\node [below of = c3, xshift=15pt] (c31) {Routing Algorithms};
\node [below of = c31] (c32) {Routing Designs};
\end{scope}
\foreach \value in {1,2,3,4}
  \draw[->] (c1.195) |- (c1\value.west);
\foreach \value in {1,2,3,4}
  \draw[->] (c2.195) |- (c2\value.west);
\foreach \value in {1,2}
  \draw[->] (c3.195) |- (c3\value.west);
\end{tikzpicture}
\caption{Classification of Papers}
    \label{figClassification}
\end{figure}
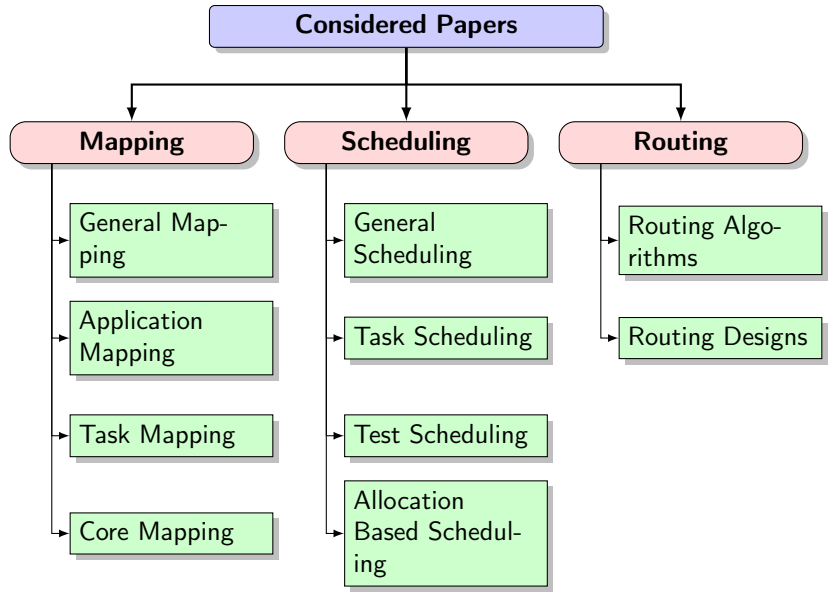

\subsubsection {Mapping Model in 3D NoC}
The 3D NoC mapping platform is composed of resource nodes, routing, and physical links. Network mapping on a chip is the process of assigning logical IP cores with known communication relations and traffic volumes on a given Task Characteristic Graph (TCG) to the corresponding resource nodes of the Architecture Characteristic Graph (ARCG) under specified system constraints, allowing various applications to be completed smoothly and efficiently under the constraints \cite{92}. 

3D NoC mapping algorithms can be divided into two categories: i) the conventional mapping optimization techniques employed in the early days and ii) new intelligent mapping algorithms based on heuristics. Traditional mapping methods primarily comprise the exhaustive technique, linear programming algorithm, and branching algorithm, and they are designed to tackle problems when the early structure of the on-chip network is basic and their application is rare. The majority of current research is focused on intelligent algorithms, which are search mechanisms designed to solve optimization problems by mimicking natural laws or biological characteristics, such as the genetic algorithm, ant colony algorithm, bat algorithm, tabu search algorithm, particle swarm algorithm, and simulated annealing algorithm. Traditional mapping methods produce generally accurate results, but as application scenarios get more complicated, the amount of calculation has become excessive. Meanwhile, it is even difficult to find the final result based on the existing computing capabilities. Therefore, there is no need to find all the solutions, as long as the intelligent algorithm that provides better and more practical solutions is immediately recognized by the majority of researchers.
Many diverse mapping strategies are provided in a range of publications that use both types of algorithms separately or in combination to get a viable answer.
We explore various mapping techniques such as Application mapping, Task mapping, and Core mapping (Figure \ref{figClassification}) and study them through an algorithmic perspective.

\subsubsection{Scheduling Model in 3D NoC}
The time spent on computation and communication to execute an application is a key component in assessing its performance. While the calculation time is primarily dictated by the architecture of the IP cores, the communication time is determined not just by the routing protocol but also by how the tasks are scheduled.  The execution order of tasks must be established to achieve optimized performance. This is referred to as the scheduling problem which can be done using a scheduling algorithm. Scheduling is the act of creating a schedule, which is a timetable for planned occurrences. The process of mapping answers the question ‘where’, but to answer ‘when’, scheduling is required. Scheduling is the time ordering of tasks and communications on their assigned resources, which assures the mutual exclusion between executions of tasks on the same resource. If multiple numbers of tasks of an application are mapped onto one core, then task scheduling is encountered. Given an application task graph mapped onto 3D NoC architecture, scheduling is the time ordering of tasks and communications determining the order in which tasks and transactions between them are to be executed such that deadlines (for real-time tasks) are met and some predefined parameters are optimized. A scheduling problem can be thought of as either a constraint satisfaction or a constrained optimization problem. We can define a scheduling problem by:\\
1. A set of time intervals defining activities, processes, or tasks to be done.\\
2. A set of temporal limitations, i.e. definitions of potential connections between the start and end time of the intervals.\\
3. A set of specialized constraints i.e. description of the complicated interactions on a set of intervals due to the state and finite capacity of resources.\\
A suitable  scheduling approach is one that satisfies design requirements such as timing constraints, data precedence constraints, and memory size constraints. It should also optimize the design goals such as reducing energy usage, improving timing performance, and balancing memory utilization. Many scholars have employed distinct strategies for the same, such as using deterministic and heuristic methods, while others have incorporated mathematical models such as Integer Linear Programming (ILP), Non-Linear Programming (NLP), and Mixed Integer Linear Programming (MILP).
We group these algorithms in four types of scheduling categories as shown in Figure \ref{figClassification}. 
(1) General Scheduling (2) Task Scheduling (3) Test Scheduling (4) Allocation-based Scheduling.

\section{Mapping and Scheduling : Algorithmic Survey} \label{AlgorithmsMapSched}

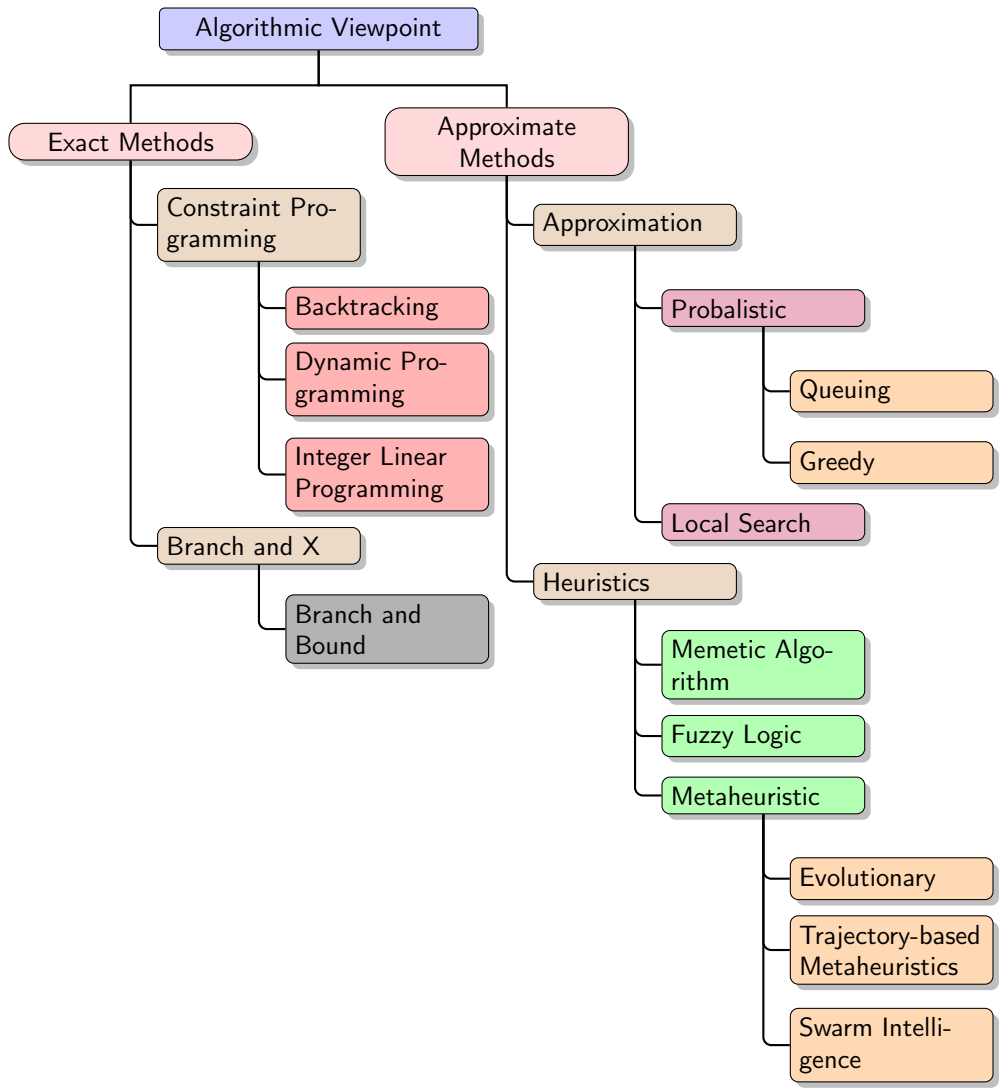
\begin{figure}
\centering
 \begin{tikzpicture}[
      basic/.style  = {draw, text width=4cm, drop shadow, font=\sffamily, rectangle},
            ROOT/.style = {basic, rounded corners=2pt, thin, align=center, fill=blue!20},
              L1/.style = {basic, rounded corners=3pt, thin, fill=purple!20},
              L2/.style = {basic, rounded corners=6pt, thin,align=center, fill=pink!60, text width=3cm},
              L3/.style = {basic, rounded corners=3pt, thin,align=left, fill=brown!30, text width=7em, grow=down, xshift=0.5em, anchor=west, 
      edge from parent path={(\tikzparentnode.south) |- (\tikzchildnode.west)}},
edge from parent/.style = {draw, thick},
              LD/.style = {level distance=#1ex},
             LD1/.style = {level distance=7ex},
             LD2/.style = {level distance=13ex},
             LD3/.style = {level distance=21ex},
         level 1/.style = {sibling distance=50 mm}
                        ]
\node[ROOT] {Algorithmic Viewpoint}
    [edge from parent fork down]
    child{node[L2] {Exact Methods}
      child[L3,LD1]   {node[L3]   {Constraint Programming}
          child[L3,LD1] {node[L3,fill=red!30]   {Backtracking}}
          child[L3,LD2] {node[L3,fill=red!30]   {Dynamic Programming}} 
          child[L3,LD3] {node[L3,fill=red!30]   {Integer Linear Programming}}
                        }  
      child[L3,LD=34]  {node[L3]   {Branch and X}
       child[L3,LD1] {node[L3,fill=black!30]   {Branch and Bound}}
      }
            }
    child{node[L2] {Approximate Methods}
      child[L3,LD1]  {node[L3, minimum width=0cm ]  {Approximation}
          child[L3,LD1]  {node[L3,fill=purple!30]   {Probalistic}
              child[L3,LD1]  {node[L3,fill=orange!30]   {Queuing}}
              child[L3,LD2]  {node[L3,fill=orange!30]   {Greedy}}
          }
          child[L3,LD=25]  {node[L3,fill=purple!30]   {Local Search}}
      }
      child[L3,LD=37]  {node[L3]   {Heuristics}
         child[L3,LD1]  {node[L3,fill=green!30]   {Memetic Algorithm}}
         child[L3,LD2]  {node[L3,fill=green!30]   {Fuzzy Logic}}
           child[L3,LD=18]  {node[L3,fill=green!30]   {Metaheuristic}
           child[L3,LD1]  {node[L3,fill=orange!30]   {Evolutionary}}
              child[L3,LD2]  {node[L3,fill=orange!30]   {Trajectory-based Metaheuristics}}
              child[L3,LD3]  {node[L3,fill=orange!30]   {Swarm Intelligence}}
           }
      }
            };
\end{tikzpicture}
  \caption{Classification based on Algorithms  }
  \label{figalgo}
\end{figure}

Previous works have mainly focused on mapping and scheduling techniques in design/run time or optimization parameters. However, to the best of our knowledge, none covers the mapping and scheduling methods from an algorithmic standpoint. Authors in \cite{realtimesurvey} presented a nice survey of the scheduling techniques concerning real-time tasks. Our work is structured around several algorithms as shown in Figure \ref{figalgo}. We have discussed various techniques that cover exact/deterministic methods that always give optimal solutions and approximate methods that are further divided into approximation and heuristics. All the papers that adopt the aforementioned algorithms are included in Table \ref{AlgoTree_Table}.

\begin{table*}[!ht]
\centering
\caption{Algorithms used in papers covered}
\label{AlgoTree_Table}
\normalsize
\begin{tabular}{|m{3.8cm}|m{6cm}|m{4.5cm}|}
\hline
{\textbf{Algorithm}} & {\textbf{Technique}} & {\textbf{Reference Papers}}\\ \hline
\multirow{3}{*}{Constraint Programming} &
Backtracking &
\cite{13},\cite{34},\cite{35},\cite{59} \\
\cline{2-3} &

ILP &
\cite{6},\cite{15},\cite{73},\cite{77} \\
\cline{2-3} &

Bottom-To-Top &
\cite{1_Bing2019},\cite{7} \\
\hline

Branch and X &
Branch and Bound &
\cite{4},\cite{74},\cite{78}\\
\hline

Queuing &
Highest level first &
\cite{14} \\
\cline{2-3} &
Highest Priority First &
\cite{47}\\
\cline{2-3} &
Earliest Deadline First &
\cite{28}\\
\hline

Greedy &
Balanced min k-way partitioning &
\cite{85} \\
\cline{2-3} &
Dijkstra’s algorithm &
\cite{10}\\
\hline

Local Search &
STAGE &
\cite{2},\cite{53} \\
\cline{2-3} &
Approximation &
\cite{30} \\
\hline

Evolutionary &
Genetic Algorithm & 
\begin{tabular}[c]{@{}l@{}}\cite{5},\cite{10},\cite{12},\cite{13},\cite{19},\cite{21},\\\cite{24},\cite{67},\cite{72},\cite{75},\cite{76}, \cite{83},\cite{92} \end{tabular}\\
\cline{2-3} &
Hybridised Constructive Heuristic (HyCH) &
\cite{84} \\
\hline

Swarm Intelligence & 
Bat Algorithm &
\cite{86} \\
\cline{2-3} &
Particle Swarm &
\cite{18},\cite{70},\cite{81},\cite{87},\cite{88},\cite{89},\cite{97}\\
\hline

Trajectory based MetaHeuristic &
Simulated Annealing &
\cite{72},\cite{83}.\cite{80},\cite{54} \\
\hline

Fuzzy Algorithms &
Fuzzy Logic &
\cite{8}\\
\hline
Voltage Based Algorithm &
Voltage and Frequency Scaling &
\cite{25},\cite{27} \\
\hline

Memetic Algorithm &
Linear Search &
\cite{1_Bing2019},\cite{16} \\
\hline

ML Based &
Clustering &
\cite{20},\cite{76}\\
\hline
Lottery Algorithm &
Lottery Arbiter  &
\cite{38},\cite{39},\cite{47} \\
\hline
\end{tabular}
\end{table*}

\subsection{Mapping} \label{Sec:FirstModel}
One of the most crucial parts of designing an NoC structure is mapping. It is the process of determining the arrangement of IP cores on the NoC structure.
Here, we have classified existing works based on algorithmic paradigms as shown in Figure \ref{figalgo}.

Heuristics under approximate methods (Figure \ref{figalgo}) provide a good estimate of the solution, but we do not know how accurate it will be (intuitive). They are used to solve NP problems and decrease the time complexity of problems by giving quick solutions. An evolutionary algorithm is one such heuristic-based approach used for solving problems that cannot be easily solved in polynomial time, such as classically NP-Hard problems, and anything else that would take far too long to exhaustively process. One type of evolutionary algorithm is the \emph{Genetic algorithm} (GA), which is used for mapping and scheduling in a wide range of research papers.		
GA is considered as an efficient algorithm for the mapping in NoC structures. A genetic algorithm creates an initial, random pool of viable solutions (chromosomes), which are assessed on each iteration (generation) by assigning fitness scores based on an objective function that steers the drive toward an efficient solution to the issue \cite{Ga}.
Authors in \cite{5} employ GA for the placement of application tasks on 3D NoC structure targeting three optimization types - thermal aware, communication aware, and hybrid optimization. The hybrid optimization approach, targeting an optimal trade-off between communication and thermal aware methods, reduces communication significantly which in turn reduces the power consumed by the chip. As a result, in addition to the reduction in communication, the average temperature of the chip in this approach is very close to that obtained in the initial thermal-aware method.
Moreover, since the peak temperature is also reduced, this approach allows the mapping of multiple tasks on each processing element, which gives a significant boost in performance. A power-aware mapping for allocation of IPs over mesh nodes is proposed in \cite{10} where GA is used to determine the network mapping that uses the least amount of power, while the Dijkstra algorithm is used to identify the shortest routing path. The authors introduce \( P_{link_{ij}}\) and \( P_{router_{ij}} \) to represent the power consumed by links and routers in 3D-mesh networks. Next, a power matrix \( P = [P_{ij}] \), where \( P_{ij} = P_{link_{ij}} + P_{router_{ij}} \) and  a weight matrix \( W = [w_{ij}] \) that uses Dijkstra algorithm is created to find the least cost path in a graph enabling exchange of data between all IPs. The weight matrix \( W\) also helps in deriving the accurate power model of 3D NoC. Furthermore, the objective function, total power consumption \( P_{total} \) is minimized.\\
If more than one metric is examined in any case, a multi-objective genetic algorithm provides a better result. SOGA (single-objective genetic algorithm) is a heuristic search technique that is based on natural selection and evolution \cite{soga}. Instead of (SOGA), a rank-based multi-objective genetic algorithm is applied in \cite{12} to minimize energy usage and packet latency of 3D NoC. In this method, three operators i.e selection, crossover, and mutation are defined which use real number encoding instead of traditional encoding where binary data is used as chromosomes. The selection operator is based on ranks of chromosomes instead of fitness and authors access the packet latency under two conditions: no congestion and congestion. The algorithm identifies the nondominant fronts and eliminates their nodes. Thus, it finds an approximate pareto front. However, this scheme completely neglects the temperature of 3D NoC. Another GA-based mapping approach has been presented in \cite{13} which optimizes the energy by mapping the task graph nodes onto the stacks, then the stacks are mapped onto the bus, and finally, the bus's position is modified to reduce transmission energy. Likewise, authors in \cite{21}, \cite{75} also employ a GA to reach the chip's thermal equilibrium and for the reduction in both temperature deviation and peak temperature.

A further approach to mapping problem is to formulate it using an integer linear programming (ILP)-based technique \cite{ilp} which is a type of exact method (Figure \ref{figalgo}). Hamedani \emph{et al.} \cite{6} use ILP and propose three mapping strategies to analyze thermal limitations and their influence on the temperature and performance of a 3D NoC. The first algorithm minimizes the peak temperature of the chip under the constraints of bandwidth, performance, and task assignment. However, the complexity of this mapping increases exponentially with the number of variables which is in the order of \( |PE|\times|E| \), where PE is processing elements and E is the edges of the communication task graph. Therefore this model may not be efficient for the next generation of chips.
The second algorithm proposes a mapping of tasks based on hierarchical partitioning using ILP with the aim of minimizing cuts between partitions. It takes into consideration, the constraint of bandwidth, task assignment, and power consumption. Although the first algorithm takes more information into account, the second algorithm produces final results quite rapidly.
Therefore the third algorithm called approximation mapping was proposed which uses the second algorithm for partitioning and the first for merging. It shows a significant reduction in peak temperature.\\ 
Effective run-time mapping on such 3-D NoC-based MPSoCs can be difficult since the arrival order and task graphs of the target applications are often unknown in advance, which is exacerbated further by strict energy constraints for NoC systems. Hence, authors in \cite{32} propose a run-time incremental mapping that reduces energy consumption by classifying the application as communication-centric or computation-centric. 
If it is communication-centric, edges with heavy data flow are mapped vertically. If it is computation-centric, high-power-consumption jobs are routed closer to the heat sink.\\ A summary of general mapping techniques is illustrated in Table \ref{GenMapping}
\def\seqinsert{\-}
\begin{table*}[htbp]
\begin{center}
\centering
\caption{Overview of general mapping techniques}
\label{GenMapping}
\footnotesize
\begin{tabular}{|m{.3cm}|m{2.3cm}|m{1.6cm}|m{2.3cm}|m{2.9cm}|m{4.5cm}|}
\hline
\textbf {Ref.} & \textbf{Algorithm} & \textbf{Target for optimization} & \textbf{Strategy applied} & \textbf{Objective} & \textbf{Limitation} \\
\hline
\cite{1_Bing2019} &
\begin{tabular}[c]{@{}l@{}}Search tree and\\ Defragmentation \end{tabular} &
TSV integration &
\begin{tabular}[c]{@{}l@{}}Application to\\ Core Mapping \end{tabular}&
\begin{tabular}[c]{@{}l@{}}Optimizing\\ communication latency\\ and temperature \end{tabular}&
\begin{tabular}[c]{@{}l@{}}i) Shows worst performance when\\ system is extremely fragmented or\\ when applications have no\\ communication among the tasks. \\
ii) No support for non-contiguous\\ mapping \end{tabular} \\
\hline

\cite{5} &
\begin{tabular}[c]{@{}l@{}}Genetic Algorithm \end{tabular} &
TFG and PE &
\begin{tabular}[c]{@{}l@{}}Mapping of logical\\ tasks onto a single\\ PE \end{tabular}&
\begin{tabular}[c]{@{}l@{}}i) Reduces power\\ consumption\\
ii) Reduces volume of\\ messages transferred \end{tabular}&
\begin{tabular}[c]{@{}l@{}} i)Increases the temperature of\\ system \\
ii)Storage overhead due to encoding \end{tabular}\\
\hline

\cite{6} &
ILP(Integer Linear Programming) &
\begin{tabular}[c]{@{}l@{}}Interconnects\\ and Topology \end{tabular}&
\begin{tabular}[c]{@{}l@{}}Mapping of tasks\\ onto different PEs \end{tabular} &
\begin{tabular}[c]{@{}l@{}}Optimizing peak\\ temperature of the\\ chip \end{tabular}&
The complexity of the mapping problem increases exponentially with the number of
variables
Only the peak temperature of the chip is considered. The distribution of temperature
fragmentation is neglected.\\
\hline

\cite{10} &
\begin{tabular}[c]{@{}l@{}} Genetic Algorithm \end{tabular}&
Topology and IP cores &
\begin{tabular}[c]{@{}l@{}}Mesh network\\ mapping \end{tabular} &
\begin{tabular}[c]{@{}l@{}}Optimizing power\\ consumption of the\\ chip \end{tabular}&
\begin{tabular}[c]{@{}l@{}}i) Very high Computational\\ complexity \\
ii) Communication latency and\\ temperature of the chip are neglected
\end{tabular}\\
\hline

\cite{12}&
\begin{tabular}[c]{@{}l@{}}Rank-Based Multi\\ Objective Genetic\\ Algorithm \end{tabular}&
\begin{tabular}[c]{@{}l@{}}Routers,\\ IP Cores \end{tabular} &
\begin{tabular}[c]{@{}l@{}}Mesh network\\ mapping \end{tabular} &
\begin{tabular}[c]{@{}l@{}}Optimizing the\\ packet latency \end{tabular}  &
\begin{tabular}[c]{@{}l@{}}i) The strong elitism may incur high\\ selection pressure, which leads to\\ premature convergence. \\
ii) They support mapping only for\\ single application systems, ignoring \\the capacity of 3D-MPSoC to handle\\ several applications during the\\ execution time.\end{tabular}
\\
\hline

\cite{13} &
\begin{tabular}[c]{@{}l@{}}Backtracking and\\ Genetic algorithm \end{tabular}&
TSV and Flits &
Mapping IP cores on buses &
\begin{tabular}[c]{@{}l@{}}Optimising the\\ Communication\\ energy \end{tabular}&
\begin{tabular}[c]{@{}l@{}}i)Execution time increases due to\\ filtering of bad results\\
ii)Other constraints like latency and\\ temperature of mapping application\\ onto the chip are neglected.\end{tabular}\\
\hline

\cite{18} &
\begin{tabular}[c]{@{}l@{}}Octahedral\\ traversal Attractive\\- Repulsive PSO \end{tabular}&
IP cores &
Mapping of IP cores onto tiles &
\begin{tabular}[c]{@{}l@{}}Optimising\\ Communication\\ energy and Latency
\end{tabular}&
\begin{tabular}[c]{@{}l@{}}i)Temperature of the cores is not\\ considered\\
ii) Designed only for a specific\\ topology \end{tabular}\\
\hline

\cite{34} &
\begin{tabular}[c]{@{}l@{}}Constraint\\ Programming \end{tabular}&
IP Cores &
\begin{tabular}[c]{@{}l@{}}Mapping of\\ Heterogeneous\\ CPU cores on chip \end{tabular}&
\begin{tabular}[c]{@{}l@{}}Limiting temperature\\ of all processing\\ nodes to\\ predetermined\\ constraints \end{tabular}&
3D mesh structures may yield shorter task completion times, without compromising thermal constraints \\
\hline

\cite{80} &
\begin{tabular}[c]{@{}l@{}}Simulated\\ Allocation (SAL) \end{tabular} &
\begin{tabular}[c]{@{}l@{}} EM-induced\\ TSV, Routers \end{tabular} &
\begin{tabular}[c]{@{}l@{}}Core to tile\\ mapping \end{tabular}&
\begin{tabular}[c]{@{}l@{}}Optimizing the\\ reliability of TSVs in\\ the 3D chip \end{tabular} &
Does not satisfy the energy along with reliability requirements \\
\hline

\cite{83} &
\begin{tabular}[c]{@{}l@{}}Cataclysm Genetic-\\based Simulated\\ Annealing (CGSA) \end{tabular}&
\begin{tabular}[c]{@{}l@{}}IP cores,\\ Topology \end{tabular}&
Application to core mapping &
\begin{tabular}[c]{@{}l@{}}Finding the best IP\\ core mapping solution\\ along with the best\\ matched 3D NoC\\ topology \end{tabular} &
\begin{tabular}[c]{@{}l@{}}i) Fabrication Cost overhead \\ ii) Premature stagnation \end{tabular} \\

\hline

\end{tabular}
\end{center}
\end{table*}
. 

Mapping is further divided into application mapping, core mapping, and task mapping. Application mapping is a process of assigning tasks to the IP cores, whereas core mapping is the process of allocating IP cores to optical routers for a particular communication purpose. In task mapping, the tasks are organized in accordance with a criterion before being mapped into IP cores.
We have summarized the algorithms used in different types of mapping as shown in Table \ref{Mapping_Types}
\def\seqinsert{\-}
\begin{table*}[htbp]
\begin{center}
\centering
\caption{Classification of algorithms based on Mapping Types}
\label{Mapping_Types}
\scriptsize
\resizebox{\textwidth}{!}
{
\begin{tabular}{|m{1.1cm}|m{0.3cm}|m{2.1cm}|m{1.8cm}|m{2.7cm}|m{3.7cm}|m{3.7cm}|}
\hline
\textbf{Mapping Type} & \textbf {Ref.} & \textbf{Algorithm} & \textbf{Target for optimization} & \textbf{Strategy applied} & \textbf{Objective} & \textbf{Limitation} \\
\hline
\multirow{5}{*}{Application} &
\cite{4} &
Branch and Bound &
\begin{tabular}[c]{@{}l@{}}Cores and Routing\\ paths \end{tabular} &
\begin{tabular}[c]{@{}l@{}} Application to core\\ Mapping\end{tabular}  & 
\begin{tabular}[c]{@{}l@{}}Reducing dynamic\\ communication energy \\consumption \end{tabular} & 
\begin{tabular}[c]{@{}l@{}}i) Only limited to core mapping\\ in regular NoC \\
ii) High computational\\ complexity when number of\\ tasks increase \end{tabular}\\
\cline{2-7} &

\cite{16} &
\begin{tabular}[c]{@{}l@{}}Memetic Algorithm\\ and Linear Search \end{tabular} &
Topology &
\begin{tabular}[c]{@{}l@{}}Mapping of IP blocks\\ to 3D NoC tiles \end{tabular} &
\begin{tabular}[c]{@{}l@{}}Analysing the efficient\\ topology for mapping based on\\ multiple objectives (total power\\ consumption of the chip,total\\ area of the chip,total delay in\\ the chip)\end{tabular} &
No consideration of Network conditions 
\\\cline{2-7} &

\cite{37} &
ARTEMIS &
\begin{tabular}[c]{@{}l@{}}Mapping Design \end{tabular}&
\begin{tabular}[c]{@{}l@{}}Runtime framework\\ for efficient dynamic\\ application mapping \end{tabular}&
\begin{tabular}[c]{@{}l@{}}Enhance the performance and\\ lifetime of 3D NoC-based chip\\ multiprocessor \end{tabular} &
Does not consider the temperature of system
\\\cline{2-7} &

\cite{81} &
\begin{tabular}[c]{@{}l@{}}Multi Objective\\ Particle Swarm\\ Optimization\\ (MOPSO)\end{tabular} &
\begin{tabular}[c]{@{}l@{}} IP Blocks and\\ Topology \end{tabular} &
\begin{tabular}[c]{@{}l@{}}Mapping to\\ application on\\ different topologies \end{tabular} &
Selection of the most suited IP block &
\begin{tabular}[c]{@{}l@{}}Does not support fault\\ tolerance \end{tabular}
\\\cline{2-7} &

\cite{68} &
\begin{tabular}[c]{@{}l@{}}Multi-Objective\\ Immune algorithm \end{tabular} &
IP cores &
\begin{tabular}[c]{@{}l@{}} Application to\\ core Mapping\end{tabular} &
\begin{tabular}[c]{@{}l@{}}Determining the best\\ mappings and optimising\\ latency and power \end{tabular} &
No consideration of other performance models like area, delay and reliability \\

\hline 
\multirow{5}{*}{Task} & 
\cite{8} & 
Fuzzy logic & 
IP cores &
\begin{tabular}[c]{@{}l@{}}Task to core\\ mapping\end{tabular}  & 
\begin{tabular}[c]{@{}l@{}}Optimizing the temperature,\\ communication delay and\\ power consumption of the\\ chip \end{tabular} & 
\begin{tabular}[c]{@{}l@{}}i) Does not fully exploit the\\ vertical links to optimize latency\\
ii) Network latency of the chip\\ is ignored  
\end{tabular}
\\ \cline{2-7} &

\cite{3} &
Partitioning and Categorization &
TSV &
High volume task to low volume task mapping &
\begin{tabular}[c]{@{}l@{}}Optimization of network\\ latency \end{tabular} &
\begin{tabular}[c]{@{}l@{}}i) Does not consider thermal\\ aspects.\\
ii) All applications reuse the\\ same NoC platform in distinct\\ time slots. This results in an\\ overhead caused by\\ reconfiguring the NoC and \\loading new applications.\end{tabular}\\ 
\cline{2-7} &

\cite{19} & 
Genetic algorithm & 
IP cores & 
\begin{tabular}[c]{@{}l@{}}Task to core\\ mapping\end{tabular} & 
Optimize  communication energy and peak temperature  &
No consideration of latency and power optimzation  \\ \cline{2-7} &

\cite{24} & 
\begin{tabular}[c]{@{}l@{}}Exhaustive\\ exploration, genetic\\ algorithms \end{tabular} & 
IP cores & 
\begin{tabular}[c]{@{}l@{}}Task to core\\ mapping\end{tabular} & 
Optimising the peak temperature and performance  & 
\begin{tabular}[c]{@{}l@{}}Does not consider\\ fragmentation which leads to\\ high communication latency\\ after running many applications.\end{tabular}\\ 
\cline{2-7} &

\cite{25} &
DVFS  & 
3D Multi cores & 
Task to core Mapping & 
Optimising temperature of the chip & 
\begin{tabular}[c]{@{}l@{}}Communication delay and \\Network latency are not\\ considered \end{tabular}
\\ 
\cline{2-7} &

\cite{84}& 
\begin{tabular}[c]{@{}l@{}}Hybridised\\ Constructive\\ Heuristic (HyCH) \end{tabular}& 
TSV & 
\begin{tabular}[c]{@{}l@{}}CTG mapped on\\ the PCTG (Physical\\ Communication\\ Topology Graph.) \end{tabular}& 
\begin{tabular}[c]{@{}l@{}}Improve Performance i.e- less \\ communication overhead,\\ execution time \end{tabular}  & 
i) Focus only on the regular topologies
ii) Temperature of the system is ignored  
\\ \cline{2-7} & 

\cite{85} & 
\begin{tabular}[c]{@{}l@{}}Balanced min\\ K-way partitioning, \\
Heuristic Algorithm\end{tabular} & 
TSV  & 
\begin{tabular}[c]{@{}l@{}}Task to core\\ mapping \end{tabular}&
\begin{tabular}[c]{@{}l@{}}Reducing communication\\ energy \end{tabular} &
\begin{tabular}[c]{@{}l@{}}No consideration of\\ temperature of system \end{tabular} \\ 
\hline

\multirow{2}{*}{Core} & 
\cite{11} & 
Dual  & 
3D routers & 
\begin{tabular}[c]{@{}l@{}}Application to core\\ Mapping and IP cores\\ to tile mapping  \end{tabular} & 
Optimising the average packet latency and communication energy & 
Increase in  average hop count and delay 
\\ 
\cline{2-7}& 

\cite{70}& 
\begin{tabular}[c]{@{}l@{}}Particle Swarm\\ Optimisation \end{tabular} & 
TSV  & 
Mapping of cores to routers & 
\begin{tabular}[c]{@{}l@{}}Decreasing the\\ communication cost \end{tabular}  &
\begin{tabular}[c]{@{}l@{}}i)Higher runtime for realistic\\ use cases.\\ 
ii) Does not consider TSV cost\\ as an optimization objective \\
iii) Network status is ignored  \end{tabular} \\
\hline

\end{tabular}
}
\end{center}
\end{table*}

\subsubsection{Application Mapping}
Application mapping has evolved into a critical component of 3D NoC /3D SoC architecture design. It performs by assigning application-related tasks to specific cores in the first instance, and then an efficient mapping technique is adopted to calculate the optimal arrangement of these cores on the nodes of NoC. The application mapping approaches must be implemented keeping in mind the constraints of many-core on-chip systems, such as bandwidth, communication time, latency, throughput, power, and energy.
Authors in \cite{1_Bing2019} introduce a three-step mapping algorithm where initially a 3D cuboid core region of a specific shape is selected for each application. Secondly, the exact locations of the core regions in the chip are determined, followed by a task-to-core mapping. Finally, during the task running, the defragmentation process is executed. Wadhwani \emph{et al.} \cite{4} present a Branch-and-Bound heuristic for smart application to core mapping in 3D NoC architecture.

We know that an application is implemented on a set of collaborative (IPs). The selection of the best-suited block from a library of IPs, as well as their physical mapping into the 3D NoC architecture, are both NP-complete problems. In this case, multi-objective solutions are preferred. Authors in \cite{82} introduce Differential Evolution to deal with the blocks mapping problem in order to implement efficiently a given application which is then broadened to multi-objective optimization in order to decrease the final implementation's hardware space, execution time, and power consumption.
Bougherara \emph{et al.} \cite{81} also uses Multi-Objective Particle Swarm Optimization (MOPSO) to produce the optimal IP and physical mapping of a particular application on three-dimensional topologies.
The majority of research papers are concerned with the mesh-based topology of 3D NoC. CastNet3D \cite{77} is a network for application mapping onto mesh-based 3D-NoCs. The study proposes an ILP formulation as well as an innovative heuristic method with the goal of minimizing energy consumption.
However, in \cite{16} in addition to mesh topology, other topologies such as torus, ring, and Butterfly Fat Tree (BFT) are used for effective mapping. They used three models to analyze the effective topology: power, area, and delay. A Knowledge-based Memetic Algorithm (KBMA) is also developed to achieve an appropriate trade-off between the cost function and the design of the 3D Noc.
 
\subsubsection{Task Mapping}
The assignment of tasks to the available IP cores is called Task Mapping. In 3D NoC, each node is connected to a router through a network interface. Hence, the task mapping stage can simply be augmented to decide on the particular core to
be used. As connections between cores in a traditional NoC are accomplished by routing messages hop-by-hop from the source to the destination, a multi-hop traversal is enabled by leveraging efficient router-bypass methods. Router data paths, in particular, may be set statically as well as dynamically to enable multi-hop traversal, allowing flits to completely avoid intermediary router pipelines, resulting in ultra-low latency performance. The performance gains can only be completely achieved if there is no contention among flows that share common connections along their routing pathways. When there is conflict, bypass pathways must be terminated early, and the corresponding flits must be halted and buffered at intermediate routers for arbitration, degrading to hop-by-hop communication in the worst-case scenario. So, authors in \cite{94} address this contention problem by introducing \emph{SMT (Satisfiability Modulo Theories)}-based contention-free task mapping and scheduling framework where an application task graph can be statically compiled to a multi-core or parallel processing platform for non-preemptive execution based on 2D and 3D.
Manufacturers are now lowering the amount of on-chip TSVs or employing 3D mesh-based NoCs with partially-filled TSVs to boost yield \cite{tsv}. However, 3D NoCs with partially filled TSVs have irregular topology. Ziaeeziabari \emph{et al.} \cite{3} propose a mapping algorithm for irregular topologies and introduce the terminology such as application graph, topology graph, and ranking parameter to answer the problem of task mapping in 3D NoC. The application graph deals with the set of tasks and the communication between them whereas the topology graph deals with the set of cores and channels connecting the cores. The ranking parameter for each task in the application graph can be expressed by the equation,
\begin{equation}
Ranking(t_{i}) = \sum_{	\forall j = 1,2,3,..|T|, i\neq j} v(d_{i,j}) + v(d_{j,i}), 
\end{equation}
where \(v(d_{i,j})\), \(v(d_{j,i})\) are volumes of communication tasks.
The algorithm has four phases 1) Partitioning of application graph into high and low volume communication 2) Mapping multi-task partitions of high-volume communications 3) Mapping multi-task partitions of low-volume communication  4) Mapping single-task partitions.\\
Another set of works considers the clustering of tasks for mapping. Clustering is a method of grouping a collection of tasks so that tasks in the same group are more related to those in other groups. A cluster-based thermal-aware task allocation technique is presented by \cite{20} which maps the tasks of an application into clusters to optimize the temperature. However, as the number of tasks increases, so does the complexity of the algorithm. So, genetic algorithms are introduced for efficient task mapping. Shen \emph{et al.} \cite{19} propose a GA for thermal-aware task mapping on 3D NoCs. The suggested task mapping technique is known as 3D-TTM, and it is divided into two stages: communication-aware group mapping and thermal-aware task scheduling. Zhu \emph{et al.} \cite{24} define a similar technique with two stages: communication-aware group (CAG) and thermal-aware scheduling (TAS). TAS is implemented at run-time, whereas CAG is implemented at design-time. The optimal mapping is selected and super tasks are developed using GA or exhaustive exploration during the CAG stage, meanwhile, the super tasks created in the CAG stage will be mapped to the cores in different layers during the TAS step.

Dynamic voltage and frequency scaling has recently been used to address heat issues on 3-D NoC architectures. By decreasing core operating voltages, DVFS can efficiently manage the temperature of 3-D MCPs. Voltage scaling \cite{25},\cite{27} has also been used for mapping in 3D NoC architectures. Thermal-aware mapping and voltage scaling(TAMVES) is proposed where task-to-core mapping is followed by voltage scaling \cite{27}, whereas authors in \cite{25} propose a two-stage approach that combines design-time mapping and DVFS with run-time thermal optimization.
Another important decision to be taken for task mapping is, among multiple cores which core is the best candidate that satisfies all the requirements. In such a situation fuzzy logic may be used that acts like human decision-making. Hence, Mosayyebzadeh \emph{et al.} \cite{8} employs fuzzy logic and maps the tasks on the basis of the linguistic rules having linguistic variables such as heat transfer, distance from the source core, and distance from the hotspot.

\subsubsection{Core Mapping}
Core mapping is one of the most challenging problems in the world of 3D NoC which is the placement of cores within the optimal accessible space of the 3D chip in the context of application-specific 3D Network-on-Chip systems. Authors in \cite{34} intend to tackle this problem with a single-stage constraint programming (CP) technique. Given a Communication Task Graph (CTG) and subsequent task allocations for the cores, heterogeneous CPU cores are assigned to the best available locations on the chip to reduce total communication costs across cores. Concurrently, the application scheduling stage is executed to select the best core types from a set of technological possibilities and to reduce the makespan, or time required to perform all compute jobs on CTG. Guo \emph{et al.} \cite{83} solve IP-core mapping and topology selection problems using Cataclysm Genetic-based Simulated Annealing (CGSA) considering 3D Optical Network-on-Chip (ONoC). In CGSA, the GA is integrated with an enhanced simulated annealing algorithm and cataclysm strategies to accelerate the searching process. In addition, to improve network reliability, CGSA is coupled with topology selection, i.e., CGSA generates the ideal mapping solution based on the best-matched 3D ONoC topology.

3D NoCs with heterogeneous architecture integrates 3D and 2D routers. However, due to the restricted amount of 3D routers and vertical links, mapping applications in 3D NoCs is difficult. So, Agyeman \emph{et al.} \cite{11} propose a technique of mapping applications to tiles such that the total communication energy is minimized for which the dimensions of NoC are determined. It was observed that symmetric NoCs have low energy consumption, average packet latency, and average hop count. Clusters are formed by grouping tasks with close communication dependencies such that each cluster contains at least one task with high communication bandwidth so as to balance traffic and reduce hotspots. Two lists are used - \( ClusterMasters \) containing tasks with high communication volume and \(ClusterSlaves\) - for mapping the tasks at vertices and their neighboring tiles respectively.
As we know, TSVs occupy a significant area and do not shrink at the same rate as gates, therefore, Manna \emph{et al.} \cite{70} propose a TSV placement to reduce the TSV count. They use \emph{Particle Swarm Optimisation}(PSO) to map routers on a partially vertically-connected 3D mesh NoC with cores of different sizes. The assignment of the routers to the vertical links is determined by particle structure and fitness function which is based on the shortest communication distance. 

\subsection{Scheduling}\label{Sec:SecondModel}

Scheduling is a mechanism that helps the device to arrange a certain factors like tasks or traffic in a particular order for the efficient performance of the device. Our device is 3D-NoC. round-robin (RR), first-Come First-Served (FCFS), GA, bottom-to-top, highest level first with estimated time(HLFET), and Exhaustive exploration are some of the most widely used scheduling algorithms in 3D NoCs.\\
A summary of scheduling techniques used in 3D NoC and its types is illustrated in Table \ref{Scheduling}.

\def\seqinsert{\-}
\begin{table*}[htbp]
\begin{center}
\centering
\caption{Overview of Scheduling Techniques}
\label{Scheduling}
\footnotesize
\begin{tabular}{|m{1.3cm}|m{0.4cm}|m{2cm}|m{1.6cm}|m{1.5cm}|m{2.8cm}|m{3.9cm}|}
\hline
\textbf{Scheduling Type} & \textbf {Ref.} & \textbf{Algorithm} & \textbf{Target for optimization} & \textbf{Strategy applied} & \textbf{Objective} & \textbf{Limitation} \\
\hline

General & 
\cite{47}  & 
Round-Robin and Highest Priority First Scheduling Algorithms & 
\begin{tabular}[c]{@{}l@{}}Router packets \end{tabular} & 
Decentralized scheduler & 
\begin{tabular}[c]{@{}l@{}}Resolving conflicts\\ between traffic flows,\\ sharing one or more \\router output ports \end{tabular}&  
\begin{tabular}[c]{@{}l@{}}Difficult to find a correct time\\ quantum during run-time\\ scheduling \end{tabular}\\ \hline

\multirow{8}{*}{Task} & 
\cite{7} & 
Bottom-To-Top & 
Run-time scheduling of task graphs & 
Task to core scheduling   & 
\begin{tabular}[c]{@{}l@{}}i) Optimizing the peak\\ temperature\\
ii) Reducing execution\\ time of applications \end{tabular}& 
Communication distance is ignored, which might lead to a higher network latency    \\ 
\cline{2-7} & 

\cite{14}& 
Highest Level First with Estimated Time(HLFET)  & 
Processors of 3D-NoC  & 
\begin{tabular}[c]{@{}l@{}}Scheduling of\\ application \\within cluster\\ of processors   \end{tabular}  & 
Eliminating the thermal hotspots in the chip  & 
Network latency of the chip is not considered     \\
\cline{2-7}  & 

\cite{24}& 
\begin{tabular}[c]{@{}l@{}}Exhaustive\\ exploration,\\Genetic\\ algorithms \end{tabular}    & 
IP cores   & 
\begin{tabular}[c]{@{}l@{}}Tasks to core \\ Mapping in a\\ scheduling \\ interval\end{tabular} & 
\begin{tabular}[c]{@{}l@{}}Optimizing\\ Communication latency\\ and peak temperature of\\ 3D system\end{tabular}& 
Numerous thermal simulations for task allocation significantly reduce efficiency \\ 
\cline{2-7} &

\cite{25}   & 
DVFS  & 
\begin{tabular}[c]{@{}l@{}}3D Multi\\ cores  \end{tabular} & 
Task to core mapping and scheduling   & 
\begin{tabular}[c]{@{}l@{}}Optimizing the peak\\ temperature of the\\ system \end{tabular} & 
\begin{tabular}[c]{@{}l@{}}i) Swapping overhead due to\\ swapping of tasks\\
ii) Only considers the power\\ balance of processors ignoring\\ the thermal balance \end{tabular}
    \\ \cline{2-7}  & 

\cite{27}  & 
Voltage scaling  & 
\begin{tabular}[c]{@{}l@{}}3D Multi\\ cores  \end{tabular} & 
Task to core mapping and scheduling   &
Optimizing throughput and energy consumption while satisfying the thermal constraints & 
i) Computation-intensive which leads to large online overhead
ii) Slow and Execution time increases by Layer-By-Layer Mapping   \\
\cline{2-7} & 

\cite{28}   & 
\begin{tabular}[c]{@{}l@{}}Earliest deadline\\ first \end{tabular}  & 
Multicore chips & 
Scheduling of tasks onto layers    & 
Reducing the heat of the chip  & 
Qualitatively allocated hot tasks to cores that are near to the heat sink rather than quantitatively estimate the thermal loads of processors   \\ 
\cline{2-7} & 

\cite{29}   & 
Rotation scheduling & 
IP multicores & 
Task to core scheduling   & 
Optimizing the peak temperature of the chip   & 
Concentration only on reducing the temperature of the top layer \\ \cline{2-7} & 

\cite{22} & 
Approximation and probability & 
\begin{tabular}[c]{@{}l@{}}3D Multi\\ cores  \end{tabular} & 
Task to core mapping and scheduling  & 
\begin{tabular}[c]{@{}l@{}}To optimize\\ temperature at a low\\ performance cost \end{tabular} & 
\begin{tabular}[c]{@{}l@{}}  i)No consideration of leakage/\\temperature dependency in\\ the analysis.\\
ii)No consideration of data\\ dependencies in an application  \end{tabular}
\\ \hline

\multirow{2}{*}{Test} & 
\cite{58}   & 
Design-for-testability (DFT)  & 
Packets and routers  & 
Test Packet Delivery to IP core & 
Optimising the chip temperature during testing & 
The scan architecture cannot provide pattern-independent low-power testing. \\
\cline{2-7} &          

\cite{61} &
Unicast-Based Multicast &
IP cores &
Test Packet Delivery to IP cores &
\begin{tabular}[c]{@{}l@{}} i)Optimizing the peak\\ temperature of chip\\
ii) Reducing test time \end{tabular} &
Disregards Network failures when delivering test packets

\\ \hline

\end{tabular}
\end{center}
\end{table*}

The requirement for scheduling techniques is driven by the need for MPSoCs to support real-time applications. Network topology, router architecture, and routing algorithms are three major operations to consider while designing a 3D NoC. Applications demand proper predictability of execution and communication times, and the router must efficiently use the limited bandwidth of the links to transmit communication data between processing elements while meeting the various criteria for each service level. Hence, a decentralized scheduler design is proposed by \cite{47} that resolves conflicts between traffic flows sharing one or more router output ports. It supports Round Robin and Highest Priority First scheduling algorithms that are utilized for best-effort and guaranteed service traffic flows to resolve conflicts between router input ports and affect the network performance.
Furthermore, we have divided the existing works into the following categories based on the factors they considered.

\subsubsection{Task Scheduling}

When numerous tasks are given to a single processor, the process of selecting the appropriate or high-priority work is referred to as task scheduling. Various scheduling strategies are mentioned in this division, which simplifies certain challenges encountered by an NoC and also aids in performance enhancement.

Many applications with high-performance requirements are modeled as task graphs.
The high computation complexity of these task graphs limits their use in run-time situations.
Different works are concerned with the placement and sequencing of tasks during application execution \cite{7},\cite{25},\cite{28}. They introduce techniques that are used to handle dynamic workload variation in applications at run-time.
Zhang \emph{et al.} \cite{7} propose a Bottom to Top scheme which is designed for run-time scheduling of task graphs. It is an efficient task graph scheduling algorithm that initially maps the task graph to the bottom layer of the 3D-NoC and then it adjusts selectively a few tasks to the top layer to reduce execution time while also maintaining low operating temperature. The thermal management process is complex as there are heat effects of vertically stacked cores and there is difficulty in balancing performance requirements and overheating. Tsai \emph{et al.} \cite{28} introduce a dynamic thermal-management framework under a modern OS that consists of two parts: Task partitioning and Scheduling. Task partitioning consists of a power-aware partition algorithm where the tasks are sorted in descending order of the peak power and then the tasks with higher peak power are assigned to the bottom cores. Following power-aware partitioning, the thermal predictor is used to compute the temperature of each core in the set of vertically aligned cores, which leads to estimating the peak temperature of the stack. Next, the Earliest Deadline First (EDF) rule is used for task scheduling, with cores arranged in decreasing order of power consumption, thermal resistance, and cooling quota, and task speeds sorted in decreasing order. The results inferred that overheating is prevented without the Quality of Service guarantee being violated.
Chaturvedi \emph{et al.} \cite{25} combine design time mapping and DVFS (Dynamic Voltage Frequency Scaling) with run-time thermal optimization for the optimization of the peak temperature of the system.

Task scheduling is considered an effective approach for eliminating the thermal hotspot without introducing any hardware overhead. However, centralized thermal-aware task scheduling algorithms for 3D-NoC have been limited for incurring high computational complexity as the system scales. So to overcome this, a distributed agent-based thermal-aware task scheduling algorithm for 3D-NoC is proposed in \cite{14}. The algorithm organizes the 3D-NoC processors into clusters, and application scheduling is done simultaneously within each cluster. Three types of software agents handle all schedule procedures. The authors also develop a centralized scheduling algorithm using the same heuristics as their distributed approach for comparison. The distributed algorithm achieves lower peak temperature without increasing the execution delay and the scalability of the distributed algorithm also significantly outperforms the centralized algorithm.
Some researchers offer OS-level thermal-aware techniques based on task scheduling. An effective temperature management method with OS-level thread scheduling is introduced in \cite{9} based on 3D Mesh multi-core system. A data-aware task scheduling for each thread on a cool processor is proposed. The main aim of the algorithm is to select the optimal thread to schedule. It has an advantage over conventional and random scheduling. Optimization of system performance is done using this algorithm under temperature constraints.
Another OS-level thermal prediction model for a 3D chip is proposed in \cite{29} where a task scheduling algorithm based on rotation scheduling aims to reduce the peak temperature on the chip. Rotation Scheduling to Minimize Temperature (RSMT) addresses data dependencies in applications and can generate schedules that effectively reduce peak temperatures of cores by up to 8$^{\circ}$C and thus improving the thermal management process.

\subsubsection{Test Scheduling}
With the increasing demand for embedded cores, the implementation of an efficient architecture of core communication is becoming the new bottleneck of SoC performance. A system-on-chip designer embeds cores that are not manufactured, and hence are untested \cite{test}. To increase the testing efficiency of IP core in 3D NoC, research on 3D NoC test scheduling under many constraints is being done.
A possible issue with 3D NoC-based test access is the formation of hotspots owing to stacking and the high toggling rates associated with structural test patterns utilized for industrial testing. Hotspots and high temperatures can cause the system components to fail as it is reported by Feng \textit{et al.} that with an increase in every 10$^{\circ}$C, the failure rate of any hardware component increases by double \cite{tempFailure}. This hampers the overall system performance. To eliminate hotspots, Xiang \emph{et al.} \cite{58} present a thermal-driven test scheduling strategy in which the entire NOC bandwidth is utilized to transmit test packets.
A Unicast-based multicast scheme \cite{61} has also been introduced which avoids delivering test packets to hotspots in the NoC by utilizing the homogeneity of the cores.

\subsubsection{Allocation-based scheduling}

3D NoCs are prone to exhibit severe thermal problems due to high power density
and long heat dissipation paths in the vertical directions \cite{k_putt}. Hotspots can arise quickly in 3D NoCs because of the uneven temperature distribution, resulting in a shorter chip lifespan and worse system dependability.
We have divided this category based on the allocation of jobs or tasks. So we need to balance the temperature among different cores present in 3D NoC architectures. One method for balancing the temperature distribution is to assign and migrate tasks between the cores. It not only enhances system performance while keeping the temperature limit the same, but it also consumes less hardware implementation cost. Researchers have used the terms task allocation or job allocation for various scheduling algorithms.

Adapt3D \cite{22} investigates DVFS and workload migration/scheduling techniques for 3D systems. This technique evenly distributes application loads taking the location of the cores into account and balances the rise in temperature by assigning tasks according to the thermal probabilities of cores. A value for each core is calculated using both the thermal index of each core and thermal behaviors.
The temperature history of each core is considered and a thermal index is assigned to each core to differentiate the position of the cores. If the thermal index is low, the core is less prone to hotspots, indicating that it is closer to the heatsink. The thermal probability of each core is determined, and new workloads are allocated to cores based on their probability values. These values are regularly updated at scheduled intervals using a weight factor and the thermal index. Adapt3D performs nearly as well as the default system setup (dynamic load balancing: the default task scheduling policy in modern operating systems), but significantly lowers hotspots. Furthermore, when Adapt3D is paired with DVFS, hotspots are decreased by an additional 20-40\%.

Next Liu\emph{et al.} \cite{26} optimize the temperature of the chip by allocating hotter jobs to the cores which have less thermal resistance whereas the colder jobs are allocated to cores that have high thermal resistance in order to take the heat off from the chip. This is done by sorting the jobs in the descending order of their power dissipation, then calculating the estimated temperature of the core after executing a specific job, and then the job is allocated to the core only if the estimated temperature is less than the critical temperature otherwise it will be allocated to the next core (cores will be sorted in the ascending order of their thermal resistances). If the temperature of the chip is not balanced then the jobs with low power dissipation will be allotted to the cores with small thermal resistances. 
Another thermal aware scheduling approach where cores are allocated according to temperature is introduced in \cite{30}. The authors use an approximation approach and propose an online-task allocation algorithm, where the tasks are sorted in the descending order of their powers and then an incremental update of thermal simulation is performed so that the same task is allotted to cores (approximation approach). Then the task is allocated to the core which has the lowest temperature rise. Similarly, all the tasks are allocated.\\
A thermal-aware optimization is proposed in \cite{31} for 3D MPSOCs where the first part of the proposed approach is a power balancing algorithm in which the tasks are prioritized in the order of Earliest Finish Time (EFT) and then the energy consumed by the tasks assigned to a processor is calculated and then the task is assigned to the processor which has a less thermal impact. The second part of the proposed approach is slack distribution and voltage scaling. Task slack distribution is used to permit voltage scaling, which facilitates temporal power density and peak temperature minimization. In the first step of slack distribution, the slacks of all tasks are calculated and then the slacks are grouped into slack paths, then paths are arranged in the descending order of slack, and then a slack is assigned to tasks based upon the values of task execution time, thermal weighted energy required for a task, total path energy consumption and then in the last step all task slacks are computed again and then the voltage of the task is scaled. The third part of the proposed approach is an Iterative hotspot mitigation algorithm in which the thermal hotspots are eliminated. This is explained in Algorithm \ref{IterativeAlgo}.

\begin{algorithm}
\caption{Iterative hotspot Mitigation}\label{IterativeAlgo}
\begin{algorithmic}[1]
\While {peak temperature of 3D MPSoC can be reduced}
\State Compute peak temperature and find critical task set within the region of peak temperature 
\State Adjust the slack distribution between the current task and its neighboring tasks.
\State Calculate the possible peak temperature reduction for each specified task
\For{each task}
        \State Increase slack time of the task
        \State Decrease slack time of task's parent and child tasks by half
\EndFor    
\State Recompute peak temperature and record local slack adjustment
\State Record new peak temperature and then restore initial slack distribution
\State Employ the slack adjustment with the lowest peak temperature
\EndWhile
\end{algorithmic}
\end{algorithm}

Another scheduling approach is to combine clustering with allocation. Shen \emph{et al.} \cite{20} introduce a cluster-based thermal-aware task allocation algorithm (CTTA) where every single task is considered as a cluster. The CTTA algorithm is divided into three stages: clustering stage, cluster allocation stage, and in-stack adjustment. 
The clustering stage is further divided into two steps: initial clustering and cluster combination. Firstly, computation energy, total computation energy, and average computation energy of initial clustered tasks are calculated and in the next step, the cluster with minimum computation energy is merged with the cluster which has high communication energy with it. The merging takes place if the sum of the computation energy of the two clusters is less than the average computation energy else another cluster that has a second less computation energy is selected and merged with the cluster that has less computation energy. In the initial clustering, a list of tasks that are ready to be executed according to the descending order of their task levels is recorded and the task with minimum task level is fetched and it is clustered with its best predecessor as a new cluster. If the best predecessor is not found then the task itself becomes an individual cluster. In this way, remaining all tasks are clustered. In cluster allocation, the cluster with the largest communication traffic is allocated to the center core and the next cluster which has the maximum communication traffic with the allocated cluster is allocated to the next core. Similarly, all other clusters are allocated to the clustered core nodes. In in-stack adjustment, the task with high computational energy is moved to the core with minimum computation energy to reduce the temperature of the system.

\subsection{Routing}\label{Sec:FourthModel}

\begin{figure}
\centering
  \includegraphics[scale = 0.70]{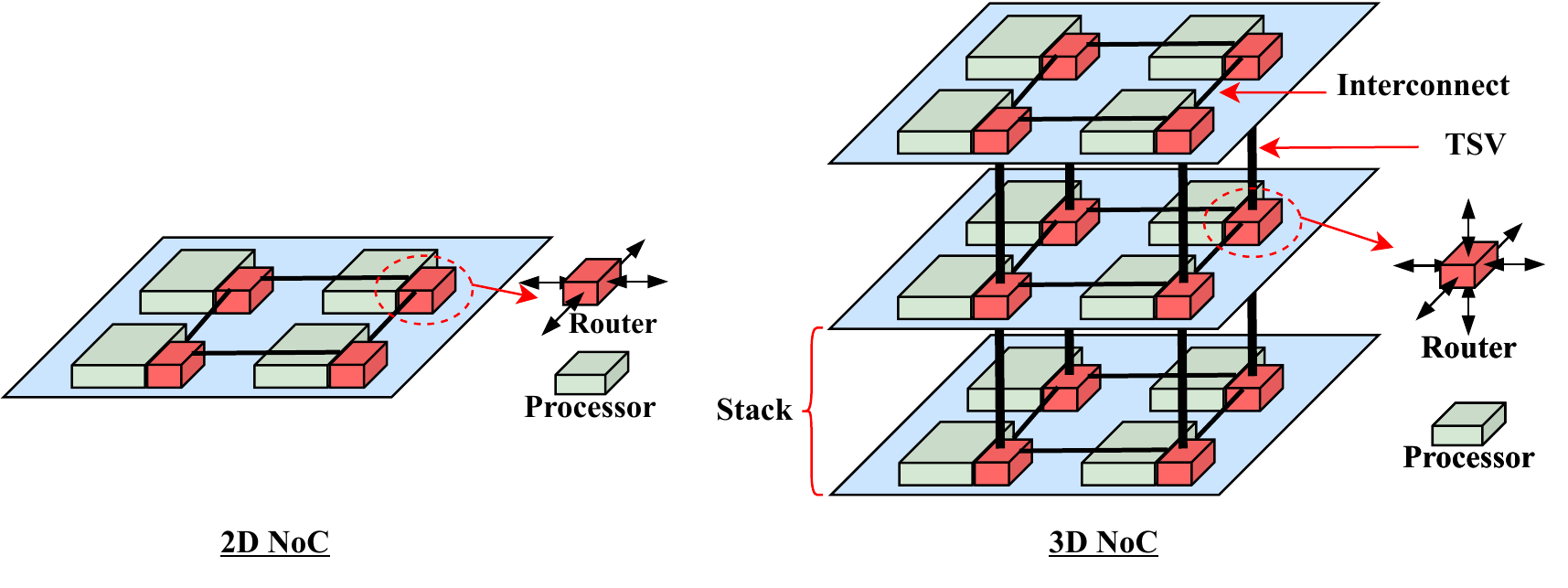}
  \caption{Implementation of router in 2D NoC and 3D Noc }
  \label{figrouter}
\end{figure}

A 3D-IC chip has many physical planes, and all IP cores may be integrated on a single physical plane or over multiple physical planes. If an NoC is utilized in a 3D-IC, it may be implemented as either a 2D-NoC (i.e., routers are implemented on a single physical layer) or a 3D-NoC (i.e., routers are implemented in multi physical plans). This is shown in Figure \ref{figrouter}. In 3D-NoC, a router on one physical plane could be connected to routers on the same plane or neighboring physical planes \cite{10}. One of the significant design tasks to consider while creating a 3D-NoC is the implementation of an efficient routing algorithm. Different routing algorithms and techniques that combine mapping with routing are present in 3D NoC. 
Look ahead XYZ routing algorithm \cite{35} is the pipeline design of the traditional XYZ routing algorithm. The virtual channels are not considered in this method for the enhancement of high traffic. This algorithm improves throughput by utilizing four pipeline systems: buffer writing (BW), routing calculation (RC), switch arbitration (SA), and the crossbar traversal stage (CT). As each hop travels through the four phases, the performance suffers and the flit latency rises. The information from the previous stage can be collected in the pipelining system, and the operations from each stage are dependent on the preceding stage. This dependence exists in the RC and SA phases, and it is eliminated in the Look ahead XYZ by having the RC and SA stages parallel.\\
Many routing algorithms are considered unsuitable for NoC applications which have
hotspots or network as priority requirements \cite{rr_network, roundr}. To address this issue, earlier algorithms included extraneous buffers into the heavy communication channels to distinguish the precedence of each channel, however extra buffers necessitate more resource space and hence consume more power. Appathurai \emph{et al.} \cite{38} suggest using a lottery arbiter in place of round robin arbiter as it can set the priority for different requests and does not require an additional buffer in the case of heavy communication. Another challenge with routing algorithms is the variable latency and throughput in each layer of a 3D interconnects. Joseph \emph{et al.} \cite{52} proposed a co-design of routing algorithms and router architectures (two routing algorithms called \emph{‘Z$^+$(XY)Z$^-$’ and ‘ZXYZ’}) to mitigate network performance due to varying throughput and latency of NoCs. Horizontal and vertical communication is modeled independently because communication within a layer is synchronous whereas communication between levels is not necessarily synchronous. The proposed models provide relevant information on their potentials and set parameters of the routing algorithms. Two routing algorithms are presented based on two principles that consider the movement of packets over multiple layers. These algorithms, when combined with router architectures, claim to have low area overhead and outperform the state-of-the-art in both theoretical and practical evaluations.

The performance of a 3D NoC and the amount of power it consumes, both are impacted by the topology of the network. BFT topology has been implemented in various routing algorithms to boost the performance of NoC architectures. A low latency energy efficient BFT-based 3D NoC design with zone-based routing strategy has been introduced in \cite{63}. It alleviates the upward traffic load problem that occurs in classical BFT especially, with a significantly large network size. Similarly, A  3D topological design of an NoC based on the BFT topology is described, together with a newly proposed table-based routing mechanism in \cite{43}. Another power-aware mapping is proposed by authors in \cite{10} that combines two algorithms. Elmiligi \emph{et al.} \cite{10} propose to use Dijkstra’s algorithm to find the shortest path routing and GA is used to find network mapping.

\section{Optimization of Attributes} \label{Attributes}

In this section, we will review the literature on various optimizations of 3D NoC. Here we consider five major categories for optimization parameters that are kept as objective while performing scheduling and mapping of applications on 3D NoC - i) Communication which consists of communication energy and communication latency, ii) Latency which consists of the packet and network latency, iii) Energy, iv) Throughput and v) Temperature.

\begin{table*}[!ht]
\centering
\begin{tabular}{|l|l|}
\hline
\multicolumn{1}{|c|}{\textbf{Attributes}} & \multicolumn{1}{c|}{\textbf{Reference Papers}} \\ \hline
Communication energy      &   \cite{4},\cite{11},\cite{17},\cite{18},
\cite{19},\cite{20},\cite{85} \\ \hline
Communication latency   & \cite{8},\cite{23},\cite{35},\cite{81} \\ \hline
Network latency              & \cite{3}  \\ \hline
Packet latency/delay                      & \cite{11},\cite{73}   \\ \hline
Latency between the links of NoC          & \cite{12}      \\ \hline
Power                & \cite{8}, \cite{9},\cite{35},\cite{36},\cite{39}, \cite{10},\cite{38},\cite{86}  \\ \hline
Energy    & \cite{13},\cite{27},\cite{32},\cite{78},\cite{79}   \\ \hline
Performance / Throughput      & \cite{9},\cite{24},\cite{33},\cite{36},
\cite{37},\cite{43},\cite{85}  \\ \hline
Overall temperature    & \cite{8},\cite{9},\cite{26},\cite{34},\cite{80}   \\ \hline
Peak temperature             & \cite{7}, \cite{8},\cite{19},\cite{20},\cite{24},
\cite{25},\cite{29},\cite{31},\cite{77} \\ \hline
Average Chip Temperature       & \cite{6},\cite{26}  \\ \hline
\end{tabular}
\caption{ Optimization attributes along with some nominal research papers }
\end{table*}

\subsection{Communication}
Data-intensive applications impose a heavy load on NoC, and communication consumes more time than computation by integrating multiple cores on the chip. Due to the increase in the volume of data and control traffic among the cores, the communication energy, and communication latency problem is addressed through various mapping techniques for communication-aware 3-D NoC designs which use different (Genetic, Heuristics, Branch or Bound) algorithms for finding optimal mapping sequence. If the whole overall communication architecture is optimized according to the traffic patterns generated by the applications under consideration, the achievable performance of a 3D NOC can be enhanced. 

\subsubsection{Communication Latency}
A crucial feature of an efficient network design is the ability to guide packets across the network in order to boost communication throughput and minimize latency.
Authors in \cite{1_Bing2019}, \cite{23} propose a runtime communication-aware mapping algorithm to optimize performance under the thermal constraint in 3D NoCs.
Wang \emph{et al.} \cite{23} introduces an algorithm to optimize latency and overall application running time by finding 3D cuboid core region and determining the location of core regions, followed by task-to-core mapping. It further extends the work in \cite{1_Bing2019} and introduces a defragmentation algorithm to reduce system fragmentation.
Bing \emph{et al.} \cite{1_Bing2019} focuses on reducing communication latency through defragmentation in three steps: \emph{(1)} For each and every application, the shape of the core region is found using search techniques and it is characterized on the basis of MD (minimal distance to the heat sink) and NOL (number of occupied layers) values. \emph{(2)} Location of the core region is determined and task-to-core mapping is done. To find the exact locations of the core regions, an algorithm is proposed which scans X and Y directions based on the given corner and starting point to find the location. Mapping tasks to cores is done on the basis of MD, NOL values of tasks, and availability of free cores. \emph{(3)} Performing defragmentation which includes determining the migration destination for the application and then finding the path to that destination. To reduce fragmentation thermal-aware defragmentation concept was proposed which focuses on migrating applications to chip corners, keeping the free cores clustered in the center. The defragmentation algorithm is triggered only when the fragmentation metric goes above the threshold fragmentation. 

\subsubsection{Communication Energy}
The energy consumed within the system is mainly because of the communication between PEs and the computation performed by PEs. Accordingly, mapping and scheduling techniques play a vital role in determining the amount of energy consumed due to communication. As the industry moves towards many-core chips, architecture designers face challenges related to communication issues, So researchers try to optimize the communication energy using various clustering methods. A novel 3D NoC architecture is proposed in \cite{17} with low-energy features consisting of two layers. Agyeman \emph{et al.} \cite{11} optimize the communication energy by determining the NOC dimensions for the application and then clustering the tasks followed by mapping the clusters onto the tiles of the 3D NoC. Similarly, authors in \cite{20} also cluster the tasks and map them onto the core stack which is denoted as a “clustered core” and then move the tasks on the clustered cores to reduce the temperature and reduce communication energy.
Other algorithms use \emph{Balanced min K-way partitioning} and \emph{Heuristic} algorithms for the same \cite{85}.
To optimize the communication energy in the chip, authors in \cite{17} propose an architecture for 3D NoC that consists of two layers where the top layer is powered by a high supply voltage  ($ V^{H}_{dd} $), while the bottom layer is driven by a low supply voltage ($ V^{L}_{dd} $) and then applications are clustered into groups and these groups are mapped onto the proposed architecture.

Grouping the tasks and executing them in batches can be another way to optimize communication energy. To reduce communication energy, authors in \cite{85} use "Balanced min K-way partitioning and Heuristic Algorithm" where the tasks in the same X, Y coordinates in a 2D plane are grouped into a partition and then communication volume between these groups is minimized.
Bhardwaj \emph{et al.} \cite{18} aims to obtain Pareto optimal IP mappings by optimizing the communication energy using \emph{Octahedral traversal Attractive - Repulsive PSO}. 
In \cite{4}, it is proposed to use a search tree to identify an optimal mapping with the least amount of communication cost, and the applications are ranked based on their communication requirements. A preference queue containing legal nodes available for branching is maintained in the order of the lowest cost. Until an optimal solution is obtained, the next unexpanded node is selected from the preference queue and the unmapped application with the highest communication requirement is mapped. Each generated child node from this expansion is inspected and pruned if its communication cost or lower limit cost is higher than the upper limit cost. The length of the preference queue, however, affects the runtime. 

\subsection{Latency}

Optimizing the latency is of paramount importance for a computing system. Different well established mapping and scheduling algorithms have been introduced in order to reduce the communication, network and packet latency \cite{1_Bing2019},\cite{3},\cite{11},\cite{12},\cite{23} in the context of 3D NoC.

\subsubsection{Packet Latency}
Packet latency is the time delay between when a packet of data goes from one place to another. To decrease average packet latency and energy consumption, tasks with high communication requirements should be mapped as closely to each other as possible.
Agyeman \emph{et al.} \cite{11} considers heterogeneous architecture to improve the performance and latency of 3D NoCs by combining 3D and 2D routers. Each layer contains tiles which are composed of processing cores and routers connected by a network interface. It introduces an efficient core mapping technique that aims to map the applications with minimum communication energy. Initially, it determines the size of NoC (3D NoCs with variable dimension sizes under uniform traffic
patterns) and then performs architecture matching. Following that, for the mapping process, application tasks are assigned to NoC regions in the order of their communication volumes, so that high-energy consumer tasks are assigned to as many 2D routers as possible while retaining a specific number of 3D routers to improve performance. Compared to existing algorithms like Onyx \cite{onyx}, CastNet \cite{tosunCastnet} and Nmap \cite{nmap}, the proposed algorithm generates 3D NoCs with low packet latency and energy consumption. 
However, the average packet latency of Branch-and-bound and the proposed algorithm is almost the same, So further research can be done on decreasing the average packet latency compared to all other algorithms.

As network congestion may have a large impact on packet delay, two different models are employed for packet latency under no congestion and congestion conditions, which use \textit{rank-based multi-objective genetic algorithm} (RMGA) \cite{12}. It encodes the different mapping alternatives in chromosomes and obtains optimal pareto-front of the mapping problem. The mapping problem is formulated as a function ($G \rightarrow T$) where G is the application characterization graph (APCG) with each vertex referring to the IP core and T is the architecture characterization graph (ARCG) with each vertex referring to a node of NoC. The map function maps the IP core to the node of NoC. The two models will determine the average packet latency of two conditions (under no congestion, under congestion) respectively. Elitism, crossover, and mutation operators are used to explore the mapping space (developing new mapping alternatives). The mapping alternatives are evaluated through an analytical model by using real number encoding for chromosomes and the selection operator depends upon the rank of chromosomes which is calculated using the following equation.

\begin{equation}
Rank(x, t) = 1 + nq(x, t)
\end{equation}
 
where \emph{nq(x,t)} refers to the number of solutions dominating solution x at generation t. In this algorithm non-dominant fronts will be identified and nodes on those fronts will be eliminated, and then the elastic selection is adopted to select the nodes. Thus, RMGA efficiently succeeds in finding an approximate pareto-front and optimizing the packet latency.

\subsubsection{Network Latency}
A mapping algorithm attempts to find a core on the chip for each task in order to reduce overall hop count, power consumption, bandwidth needs and network latency. The irregularity of 3D NoCs has a significant impact on 3D mapping techniques. This means that the mapping technique cannot take into account the same communication capabilities for all chip cores. In fact, in a 3D NoC, certain topologically near cores may have greater distance than other topologically distant cores. So, an algorithm that reduces network latency and minimizes hop count, 3D-AMAP is proposed for 3D NoCs with partially filled TSVs having irregular topologies in \cite{3}. The algorithm follows four phases of task mapping - (1) \emph{Application graph partitioning}, where average communication volume of the application graph is compared with the communication volume between tasks, and communications are categorized as High and Low volume communications. 
(2) \emph{High Volume Multi-task Partitioning}, where the task with a high ranking is first mapped on a layer having sufficient free cores followed by mapping of tasks having high communication volume with an already mapped task until all the tasks are mapped.
(3) \emph{Low Volume Multi-task Partitioning}, where task partitions follow a similar process as (2) after being sorted according to their total intra-partition communications in descending order.
(4) \emph{Single Task Mapping}, which maps tasks in the descending order of communications to an empty core having minimum hop count.
The algorithm ignores low-volume communications and reduces traffic in TSVs which in turn reduces network latency.

\subsection{Power or Energy}
With the increasing number of IP cores integrated into the processor, the power consumption of the chip continues to escalate. In 3D NoC, power consumption has a significant impact on system performance, hence, reducing power consumption has become a crucial design consideration for 3D NoC. To meet the ever-increasing communication and low-power demands, 3D NoC designs must be optimized for power and energy consumption.

\subsubsection{Power}
Chip designers lack the tools required to efficiently execute their applications at various levels of the design hierarchy. To obtain the greatest performance, a design technique for low-power 3D-NoCs applications is required. Elmiligi \emph{et al.} \cite{10} use GA to determine the optimal 3D-NoC mesh network mapping that consumes the least amount of power for a certain application. They claim to have solved the optimization problem in less than four minutes, compared to an exhaustive search that took three days and still did not identify the lowest power consumption. Dijkstra's Algorithm is used to find the shortest path routing and GA is used to find the optimum mapping with minimum power consumption for a given application. A bat algorithm (BA) is introduced in \cite{86} for the energy-aware mapping technique for 3D NoC. It requires an improved control strategy to switch between exploration and exploitation in the accurate instant
and it needs proper parameter tuning for a better search.
In an NoC, the routers consume the majority of energy, and energy consumption increases nonlinearly with the number of input ports. Karthikeyan \emph{et al.} \cite{39} uses the Lottery algorithm to reduce the power consumed by asynchronous 3D NoC by distinguishing different priorities of the input port and ensuring that it responds to the higher priority port.

\subsubsection{Energy}
The popularity of battery-driven devices brings the limitation of power availability into consideration, making energy-aware mapping an important research area. Although improvement of energy reduction techniques will facilitate continuous performance betterment, it is seldom the sole optimization criterion. This leads to a trade-off between energy consumption and other user requirements like performance and quality of the solution. 
It is known that vertical links for communication nodes in 3D NoC are shorter than horizontal ones, therefore they are quicker and use less energy. Hence, Nalci \emph{et al.} \cite{77} minimize the energy consumption of application by utilizing vertical links for communicating nodes as much as possible.
Wang \emph{et al.}\cite{32} uses dynamic energy technique to reduce the energy consumption which is generated due to the data communication between the processing cores. It classifies the application as communication centric or computation-centric after it specifies a certain condition. If the application is communication-centric then the communication edges which have high traffic volume are mapped in vertical dimensions so that the dynamic energy consumption is reduced. The authors compare the proposed algorithm to branch and bound (BNB) \cite{bnb} and two heuristic algorithms (Temperature balance (TB) \cite{tb} and Temperature-Aware Low Power Mapping (TL) \cite{tl}). The experiments show that their proposed algorithm reduces the computational and communication energy more efficiently.
Another algorithm that reduces the dynamic communication energy consumption is listed in \cite{75}. It is an energy-efficient mapping algorithm that employs the BNB strategy. 
A \emph{Traffic Equilibrium Mapping} method is proposed in \cite{13} for the energy minimization in the chip of 3D NoC-Bus mesh architecture.

Liao \emph{et al.} \cite{27} aim to optimize the energy consumption by \emph{Task-to-core} mapping which is followed by voltage scaling. It specifies three components for selecting tasks to be scaled for decreasing temperature and energy: group-power computation (GPC), heat-dissipation rate (HDR), and power-driven temperature reduction (PDTR) and satisfies the thermal constraints by calculating a \emph{TEA-FACTOR} which is used for deciding the priority of task for scaling. 
\begin{equation}
    TEA-factor(t_{i}) = GPC_{i}\times HDR_{i}\times PDTR_{i}
\end{equation}
where \(GPC_{i}\) denotes the group power of the cores containing the
running task, \(HDR_{i}\) denotes the heat-dissipation rate of the core with running task and \(PDTR_{i}\) denotes the power reduction after scaling task in the schedule.

\subsection{Throughput}
Throughput refers to the performance of tasks by a computing service or device over a specific period of time. Throughput in 3D NoC refers to the number of packets that successfully pass from source to destination paths \cite{through}. The topology of a network affects the performance and power consumption of 3D NOC. So, to get an unprecedented performance gain and optimize throughput by combining the benefits of NoC and 3D ICs, Bose \emph{et al.} \cite{43} introduce a new 3D topological NoC design which is based on the BFT topology with an efficient table based uniform routing algorithm for 3D NoC. 
Throughput can be increased only if the temperature of the system is under control. Authors in \cite{27}, \cite{30} optimize the throughput by satisfying the thermal constraints. Liao \emph{et al.} \cite{30} presents an approximated task assignment technique to enhance 3D multi-core processor by rewriting temperature equations to provide an incremental thermal update. The incremental update strategy is used to first determine the current temperature of each core. Then, among the unassigned cores, a new arrival job is assigned to the core with the lowest temperature rise. The above process terminates when all new arrival tasks are assigned to cores. Authors in \cite{27} optimize the throughput by allocating tasks with larger power to upper-layer cores which are closer to the heat sink. As cores at the layer farther from the heat sink dissipate heat more slowly, authors prevent overheating by not assigning tasks to these cores.

Techniques via Task Scheduling \cite{9},\cite{24},\cite{69} have been proposed for optimizing the performance of 3-D NOCs which also satisfy thermal constraints. Fu \emph{et al.} \cite{9} design a simulator and select the coolest thread using a temperature controller. Then, a data-oriented task scheduling for each thread on a cool processor is done and the algorithm aims to select the optimal thread to schedule. Whereas Zhu \emph{et al.} \cite{24} present a two-stage scheduling algorithm, namely the communication-aware group (CAG) stage and thermal-aware scheduling (TAS) stage which is implemented at design-time and at run-time respectively. \cite{69} also introduces a run-time thermal management technique (RTM) which uses the greater thermal coupling among the 3D NoC's vertically stacked routers. A proactive strategy is utilized to optimize performance with varying temperatures where a traffic-aware downward routing shifts horizontal routing power to the bottom layer and adjusts traffic volume to avoid packet congestion.\\
\cite{37} present a \emph{runtime aging-aware application-mapping framework} called ARTEMIS for 3D NoC-Based Chip Multiprocessor. It enhances the performance and lifetime of 3D NoC-based chip multiprocessors and manages aging in circuits and the power delivery network (PDN) simultaneously. 

\subsection{Temperature}
Although 3D ICs have many advantages like decreasing interconnect power, scaling of chip dimensions etc., it also deals with various thermal problems. 3D ICs also have high power density which increases chip temperature that further leads to performance degradation. In this case mapping can be used to control chip temperature. Hamedani \emph{et al.} \cite{6} reduce the temperature by partitioning the task graphs into subgraphs and mapping the subgraphs using ILP. Wang \emph{et al.} \cite{9} use task scheduling for optimizing the temperature and performance loss of the chip. 
Another thermal-aware task scheduling policy consisting of two stages is proposed in \cite{24} which exploits the application and system architecture characteristics to decouple the mapping of task graphs for the performance and peak temperature optimization into two stages namely communication-aware group stage and temperature-aware scheduling stage.

\subsubsection{Peak Temperature}

The heat generated in the chip is dissipated through a heat sink. The dissipation of heat generated in layers farther from the heat sink is much more difficult. Therefore novel Bottom-to-Top (B2T) task scheduling scheme is proposed in \cite{7} which aims to reduce the peak temperature of the chip by keeping power consumption in such layers lower than those closer to the heat sink. When the part of power consumed by farther layers reduces, less heat is generated in them which results in a lower peak temperature of 3D NoC.

The performance of 3D NoC can be boosted if we reduce the peak temperature as well as optimize the network competition. Since a router can be connected to more than one IP core, this leads to a network competition problem where multiple IP cores and neighbor routers may compete for the same output port of a router. A temperature and network competition-aware mapping algorithm \cite{76} is proposed to reduce the peak temperature and decrease the network competition. This algorithm realizes the multi-objective mapping and ensures a lower time complexity. In the first step of the proposed algorithm, IP cores are clustered according to the number of layers on the chip and the and network nodes connected to the same router are clustered together. The second step is to implement a GA for NoC mapping. Simulation results show that this method achieves an appropriate balance between peak temperature and network competition.
Authors in \cite{76},\cite{20} use clustering algorithms for the peak temperature optimization.

\subsubsection{Average Chip Temperature}
The increase in the number of the transistors and higher operating frequency of the processors increase the chip power consumption, which results in a higher power density in the chip, which in turn raises the chip temperature. Hamedani \emph{et al.} \cite{6} proposes three mapping algorithms and employs ILP to reduce the chip temperature where it partitions the task graphs into subgraphs and then maps the tasks after merging the subgraphs according to their communication rate.\\
Liu \emph{et al.}\cite{26} also propose three mapping algorithms and optimize the temperature of the chip by allocating hotter jobs to less thermal resistance cores and colder jobs to high thermal resistance cores. This is accomplished by sorting the jobs in descending order of their power dissipation, then calculating the estimated temperature of the core after executing a specific job and allocating the job to the core only if the estimated temperature is lower than the predefined threshold, otherwise, the job is allocated to the next core.

\section{Experimental/Evaluation Tools}\label{Experiment}
Let us now discuss the experimental setups researchers typically consider while evaluating the work on scheduling and mapping. The choice of simulators and tools is of great importance to assess and validate the results. Great attention is paid to record the performance metrics and implementing new architectures and designs for the mapping in 3D NoC architectures. This is the reason why various simulators and benchmarks in the techniques reviewed have been used to evaluate throughput, power consumption, or latency and for measuring the performance of mapping and scheduling in 3D NoCs.  

\subsection{Simulators}
\title{Pie Chart}
\begin{figure*}
\centering
\begin{tikzpicture}
\centering
\pie [rotate=180,text=legend ]{3.2/HDL-based NOC simulator, 15.9/Full system simulator,
6.3/Event-driven C++ NoC simulator,
33.3/Cycle Accurate Simulator, 
25.4/Thermal Simulator, 9.5/Access Noxim, 
6.3/Power Performance Simulators}
\end{tikzpicture}
\caption{Simulators used in various 3D mapping and Scheduling Techniques}
\label{figSimulators}
\end{figure*}
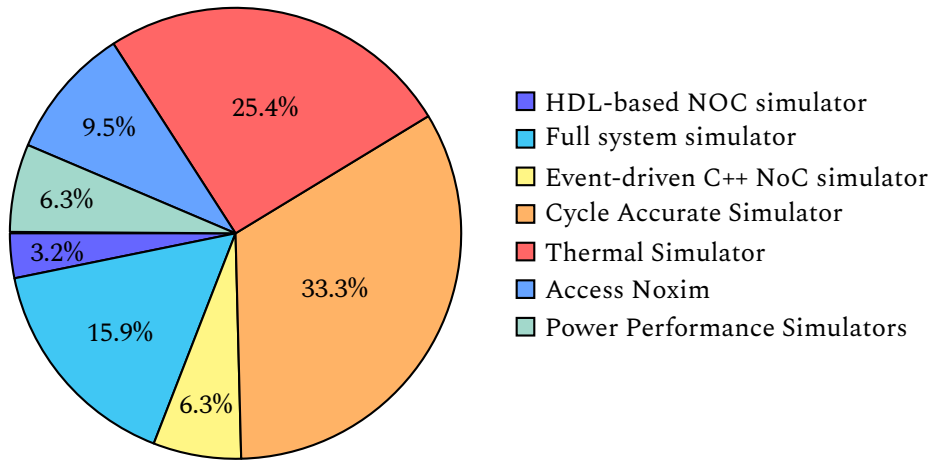

Figure \ref{figSimulators} shows different simulators used for implementing or evaluating the performances and analyzing the effectiveness of scheduling and mapping techniques in 3D SoC/NoC architectures.\\
Cycle Accurate Simulators such as Noxim\cite{noxim}, BookSim\cite{booksim}, WormSim\cite{wormsim} is used in 33\% of the considered papers, full system simulators namely Nirgam \cite{nirgam}, GEM-5\cite{gem5}, GARNET in 15\%, and Access Noxim\cite{an} constitute 10\% of the total. Thermal simulator i.e HotSpot tool \cite{hotspot} is used in 25\% of the works for temperature modeling. It is usually combined with another tool for a complete evaluation.

\subsection{Benchmarks}
\begin{figure*}
\centering
\pgfplotsset{width=8cm, compat=1.12}
\begin{tikzpicture}
\begin{axis}[
    ybar stacked,
	bar width=14pt,
    enlargelimits=0.10,
    legend style={at={(0.5,-0.5)},
      anchor=north,legend columns=-1},
    ylabel={No of papers},
    symbolic x coords={PARSEC, E3S, SPLASH, MULTIMEDIA, 
		SPEC, Based on TGFF, ALPBench, Others},
    xtick=data,
    x tick label style={rotate=45,anchor=east},
    ]
\addplot +[ybar] plot [fill=green!50,draw=black!70]  coordinates {(PARSEC,4) (E3S,8) 
  (SPLASH,4) (MULTIMEDIA,14) (SPEC,1) (Based on TGFF,5) (ALPBench,1)  (Others,7)};
\addplot+[ybar] plot [fill=yellow!90,draw=black!70] coordinates {(PARSEC,0) (E3S,0) 
  (SPLASH,0) (MULTIMEDIA,0) (SPEC,4) (Based on TGFF,0) (ALPBench,1)  (Others,3)};
\addplot+[ybar] plot [fill=red!60,draw=black!70] coordinates {(PARSEC,7) (E3S,0) 
  (SPLASH,6) (MULTIMEDIA,2) (SPEC,2) (Based on TGFF,0) (ALPBench,0)  (Others,3)};
\addplot+[ybar] plot [fill=blue!30,draw=black!70] coordinates {(PARSEC,1) (E3S,0) 
  (SPLASH,1) (MULTIMEDIA,3) (SPEC,0) (Based on TGFF,0) (ALPBench,0)  (Others,3)};
\addplot+[ybar] plot [fill=orange!60,draw=black!70] coordinates {(PARSEC,1) (E3S,0) 
  (SPLASH,1) (MULTIMEDIA,0) (SPEC,0) (Based on TGFF,0) (ALPBench,0)  (Others,1)};
  
\legend{\strut Mapping, \strut Scheduling, \strut Design, \strut Routing, \strut ML Based}
\end{axis}
\end{tikzpicture}
\caption{Benchmarks used in various 3D mapping and Scheduling Techniques}
\label{figBenchmarks}
\end{figure*}
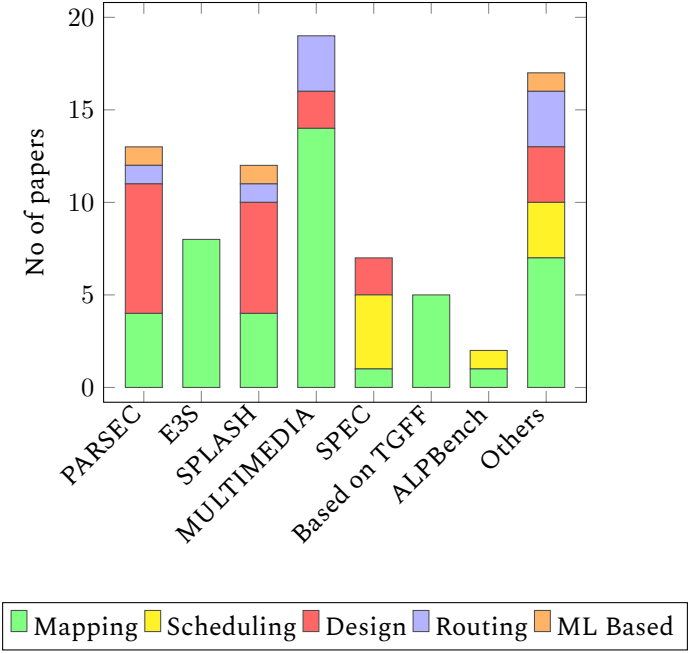

The experiments on the surveyed methodologies make use of a diverse collection of benchmark applications as shown in Figure \ref{figBenchmarks}. A considerable portion of the works we analyzed employ both synthetic and real benchmarks. 
For evaluation of mapping algorithms 31\% of total considered papers use real Multimedia benchmarks such as JPEG-encoder, MPEG 4, VOPD, MWD VOPD, DVOPD, PIP, and GSM Coder/Decoder. \emph{Embedded System Synthesis Benchmark Suite} (E3S) as automotive, consumer, networking, telecommunications and office automation constitute 18\% of mapping techniques. SPEC benchmarks such as \emph{SPEC CPU 2000} and \emph{SPECjbb} are used in 31\% of articles to evaluate most of the scheduling algorithms. 3D stacked SoCs use benchmark circuits like d695, d281, f2126, and 2f2126 for the test scheduling. 

PARSEC \cite{parsec} and SPLASH 2 \cite{splash} benchmarks have been widely used in \cite{1_Bing2019},\cite{23},\cite{37},\cite{41},\cite{44},\cite{48},
\cite{53}. PARSEC benchmarks namely, dedup, vips, fluidanimate, canneal, and BodyTrack (BT) and SPLASH-2  benchmarks, namely, fft, adix, lu, and water, were used in \cite{1_Bing2019},\cite{48} to explore and establish performance-energy-reliability trade-offs for 3D small-world NoC (SWNoC). The task graphs of the real applications are generated from the traces of SPLASH-2 and PARSEC \cite{23}.
Various benchmarks were used to assess the design of efficient NoC architectures. Some compiler benchmark suites such as MiBench \cite{Mibench} and Livermore\cite{lM} were also employed. However, PARSEC and SPLASH were the most commonly used.

\vspace{0.5cm}

\section{Conclusion and Future Scope} \label{Conclusions}
With the rising need for faster and smaller ICs, mapping and scheduling strategies play an important role in the evolution of three-dimensional NoC architectures. It is obvious that the capabilities of 3D NoC architecture can be fully leveraged by employing effective mapping and scheduling approaches. Each node in 3D is connected to a router and hence routing has a considerable impact on the efficiency of scheduling and mapping. This paper surveys various ongoing mapping and scheduling techniques through an algorithmic viewpoint for regular 3D mesh-based NoCs which satisfy various optimization constraints such as performance, communication cost, temperature, energy consumption, and reliability. We have arranged these works in this paper based on the optimization objectives, simulators and benchmarks adopted, and the approach employed for mapping and scheduling. However, there are several challenges and opportunities related to optimization, generalizability, multi-core considerations, reliability, and the application or combination of techniques with more intelligent techniques as discussed below.
\begin{itemize}
\item Although different scheduling and mapping strategies claim to optimize different parameters, most of them ignore the increasing time for optimization. Further research can be done on reducing the run time overhead of the proposed algorithms. 

The work presented by authors in \cite{37} which is a run-time application mapping framework that can be further extended to investigate support for variable process variations, and may consider a service queue model that includes the wait time of an application. Similarly, 3D mesh structures in \cite{34} can be utilized to yield shorter task completion times, without compromising thermal constraints.

\item The growth in power density and heat dissipation makes the 3D-NoC system more easily get overheated and harms the reliability of the system. The router is a potential hotspot here because of the frequent packet routing while running the applications after they are mapped. To meet the temperature criteria, the router microarchitecture and routing algorithm must be adequately developed \cite{33}. This area needs to be further explored and hotspots on NoC routers need to be considered while performing mapping, to explore further thermal optimization on 3D NoC architecture.

\item Instead of considering a single factor in scheduling objectives, we need to broaden our research to include multi-objective optimization, which involves optimizing more than one performance indicator at the same time. For instance, DVFS can be applied in \cite{24} along with the thermal-aware scheduling algorithm to further reduce the peak temperature and enhance performance. 

\item Security is a critical aspect that has not been addressed in any of the works that work on scheduling and mapping of applications. Emerging 3D systems try to incorporate a large number of layers which increases the temperature gradient because only one layer is attached to the heat sink. The temperature standard deviation for the 4-layer 3D chip is roughly 40 times greater than that of the 1-layer 2D device \cite{survey_security}. A Trojan can exploit this flaw by flooding the topmost layer with bogus and/or valid traffic, increasing the likelihood of surpassing the temperature threshold and inducing thermal throttling. Using the algorithms mentioned in this survey paper, we may include secure mapping and scheduling strategies for the upcoming research opportunities.

\item Thermal-aware test scheduling has attracted a lot of research interest as this issue has become increasingly important for high-performance testing systems. In order to handle testing scenarios in 3D NoC, researchers developed different mapping and scheduling algorithms for the 3D NoC test scheduling problem. However, an inefficient approach to the NOC testing problem does not utilize the full NOC bandwidth for testing and instead sequentially delivers tests to cores which significantly increases test time and test cost \cite{testnoc}. One solution is that we can employ partitioning of test schedules with interleaving of tests to improve the performance with respect to temperature control and total time consumed for testing the 3D NoC architectures.

\item The reliability of three-dimensional systems has yet to be considered in scheduling and mapping strategies. Future work may include a reliability model for the hybrid 2D router (as proposed in \cite{51}) to address TSV yield and failure issues, thereby increasing the quality of service. 

\item Most of the papers target mesh topology of 3D NoC while designing scheduling and mapping policy. The research on scheduling and mapping on 3D NoC can be extended for other topologies as well. Furthermore, designing policies for a generic 3D NoC topology may also be explored.

\item In recent years, the number of ML-based applications has increased dramatically. Machine Learning plays an important role in designing an efficient 3D NoC architecture. Few authors currently explore this field with some limitations \cite{50}. ML models are highly adaptive and these may be explored to find efficient mapping catering to the dynamicity of the system such as temperature, bandwidth, etc. Heterogenous multi-core architectures running multi-threaded programs result in complex task mapping and load balancing. Artificial neural network (ANN) based load balancing techniques may also be combined with the aforementioned mapping and scheduling techniques to further improve the system performance. 

\end{itemize}

\printbibliography

\end{document}